\documentclass{aa}
\usepackage{amsmath}
\usepackage{amsfonts}
\usepackage{color} 
\usepackage{graphicx}
\usepackage[varg]{txfonts}
\usepackage{natbib}
\usepackage{hyperref}
\bibpunct{(}{)}{;}{a}{}{,} 
\begin{document}
%
\title{Supernova 1987A: neutrino-driven explosions in three dimensions
   and light curves}

\author{V.~P.~Utrobin\inst{1,2} \and A.~Wongwathanarat\inst{1,3} \and
        H.-Th.~Janka\inst{1} \and E.~M\"uller\inst{1}}

\institute{
   Max-Planck-Institut f\"ur Astrophysik,
   Karl-Schwarzschild-Str. 1, 85748 Garching, Germany
\and
   State Scientific Center of the Russian Federation --
   Institute for Theoretical and Experimental Physics of
   National Research Center ``Kurchatov Institute'',
   B.~Cheremushkinskaya St. 25, 117218 Moscow, Russia
\and
   RIKEN, Astrophysical Big Bang Laboratory, 2-1 Hirosawa, Wako,
   Saitama 351-0198, Japan
}
\date{Received 13 December 2014 / accepted 8 June 2015}

\abstract{
The well-observed and well-studied type IIP Supernova 1987A (SN~1987A),
   produced by the explosion of a blue supergiant in the Large Magellanic
   Cloud, is a touchstone for the evolution of massive stars, the simulation of
   neutrino-driven explosions, and the modeling of light curves and spectra.
}{
In the framework of the neutrino-driven explosion mechanism, we study the
   dependence of explosion properties on the structure of different blue
   supergiant progenitors and compare the corresponding light curves with
   observations of SN~1987A.
}{
Three-dimensional (3D) simulations of neutrino-driven explosions are
   performed with the explicit, finite-volume, Eulerian, multifluid
   hydrodynamics code {\sc Prometheus}, using of four available presupernova
   models as initial data.
At a stage of almost homologous expansion, the hydrodynamical and composition
   variables of the 3D models are mapped to a spherically
   symmetric configuration, and the simulations are continued with the
   implicit, Lagrangian radiation-hydrodynamics code {\sc Crab} to follow
   the blast-wave evolution into the SN outburst.
}{
All of our 3D neutrino-driven explosion models, with explosion energies
   compatible with SN~1987A, produce $^{56}$Ni in rough agreement with
   the amount deduced from fitting the radioactively powered light-curve
   tail.
Two of our models (based on the same progenitor) yield maximum velocities
   of around 3000\,km\,s$^{-1}$ for the bulk of ejected $^{56}$Ni, consistent
   with observational data.
In all of our models inward mixing of hydrogen during the 3D evolution leads
   to minimum velocities of hydrogen-rich matter below 100\,km\,s$^{-1}$,
   which is in good agreement with spectral observations.
However, the explosion of only one of the considered progenitors
   reproduces the shape of the broad light curve maximum of SN~1987A
   fairly well.
}{
The considered presupernova models, 3D explosion simulations, and light-curve  
   calculations can explain the basic observational features of SN~1987A,
   except for those connected to the presupernova structure of the outer
   stellar layers.
All progenitors have presupernova radii that are too large to reproduce the narrow
   initial luminosity peak, and the structure of their outer layers is not
   suitable to match the observed light curve during the first 30--40 days.
Only one stellar model has a structure of the helium core and the He/H
   composition interface that enables sufficient outward mixing of $^{56}$Ni
   and inward mixing of hydrogen to produce a good match of the dome-like
   shape of the observed light-curve maximum, but this model falls short of
   the helium-core mass of 6\,$M_{\sun}$ inferred from the absolute
   luminosity of the presupernova star.
The lack of an adequate presupernova model for the well-studied SN~1987A is
   a real and pressing challenge for the theory of the evolution of massive stars.
}
%
\titlerunning{SN~1987A: neutrino-driven explosions and light curves}
\authorrunning{V. P. Utrobin et al.}
\maketitle

\section{Introduction}
\label{sec:intro}
%
It is well known that massive stars die as core-collapse supernovae (SNe).
Type II-plateau supernovae (SNe IIP) represent the most numerous subclass of
   core-collapse SNe.
They are characterized by a $\sim$100 day plateau in the light curve, which is
   a generic feature of the explosion of a red supergiant (RSG) star
   \citep{GIN_71, FA_77}.
The RSG nature of SNe IIP was confirmed by the detection of their progenitors
   on archival images \citep{Sma_09}.
The theory of stellar evolution also predicts that stars in the range of
   $9-25...30\,M_{\sun}$ end their life as RSGs \citep{HFWLH_03}.
In addition to the ordinary SNe IIP, there is a group of peculiar objects
   of which SN~1987A in the Large Magellanic Cloud (LMC) is the best studied.
It was identified with the explosion of the blue supergiant (BSG)
   Sanduleak $-69^{\circ}202$ and is a serious challenge for theorists.

We still have no clear answer to why its progenitor was a compact star.
Just a couple of months after the discovery of SN~1987A, \citet{Arn_87}
   and \citet{HHTW_87} showed that a metal-deficient composition similar to
   the LMC favors the formation of BSGs.
It, however, cannot explain the high nitrogen abundance that was revealed in
   the circumstellar matter of SN~1987A by an analysis of ultraviolet
   lines \citep{Cas_87, LF_96}.
In addition, either a modification of convective mixing
   through rotation-induced meridional circulation in the star during
   its evolution \citep{WHT_88}, a restricted semiconvective
   diffusion \citep{WPE_88}, both mass loss and convective mixing
   \citep{SNK_88}, or invoking mass-loss effects in a close binary system
   \citep{HM_89, PJ_89} are required to fit the observed properties of 
   Sanduleak $-69^{\circ}202$ in evolutionary calculations.

The phenomenon of core-collapse SNe is very complex and how massive stars
   explode has only been elucidated in recent decades,
   but we still do not know exactly how the explosion engine works
   \citep{WJ_05, Jan_12, JHH_12, Bur_13}.
While this problem is difficult to solve for a variety of reasons, very
   different energies and timescales inside the star just before the
   gravitational collapse of its central iron core separate the evolution
   of the iron core from that of outer layers.
These vast differences allow us to divide the whole problem
   into two: an internal problem (the gravitational collapse itself)
   and an external problem (the collapse-initiated SN outburst)
   \citep{IN_89}, and to study them independently.
We briefly present these two approaches focusing on the well-observed and
   well-studied peculiar SN~1987A.
We start with the external problem, as it directly deals with the photometric
   and spectroscopic observations.

One of the surprises associated with SN~1987A was the observational evidence for
   macroscopic mixing occurring during the explosion.
This kind of effect is required to reproduce both the smooth rising part of
   the bolometric light curve and the major broad maximum with a timescale of
   $\sim$100 days \citep{Woo_88, SN_90, Utr_93}.
The smoothness of the bolometric light curve results from the mixing of
   radioactive $^{56}$Ni out of the center, while the wide dome-like light
   curve maximum comes from mixing hydrogen-rich matter down to the center.

The extent of mixing of radioactive $^{56}$Ni and hydrogen in the ejected envelope
   is a crucial point for understanding SN~1987A.
Observational evidence for this mixing in the envelope exists not only at
   early times, but also at late times.
At early times (during 20--100 days), the H$\alpha$ profile exhibits a striking
   fine structure called ``Bochum event'' \citep{HD_87, PH_89}.
At day 410, a unique high velocity feature with a radial velocity of about
   $+3900$\,km\,s$^{-1}$ was found in the 17.9 and 26.0\,$\mu$m infrared
   emission lines of [Fe II].
It was interpreted as a fast iron clump with a mass of
   $2 \times 10^{-3}\,M_{\sun}$ \citep{HCE_90}.
Later, a similar high velocity feature was detected in the 6.6\,$\mu$ infrared
   emission line of [Ni II] at day 640 \citep{CHELH_94}. 
\citet{UCA_95} analyzed the H$\alpha$ profile at the Bochum event phase
   (days 29 and 41) and identified this high velocity feature with a fast
   $^{56}$Ni clump that was moving at an absolute velocity of
   4700\,km\,s$^{-1}$ and had a mass of $\sim$10$^{-3}\,M_{\sun}$.
As to the bulk of radioactive $^{56}$Ni, measurements of the 6.6\,$\mu$m infrared
   emission line of [Ni II] and the 17.9 and 26.0\,$\mu$m lines of [Fe II]
   at day 640 showed that it was moving with a maximum velocity of
   $\sim$3000\,km\,s$^{-1}$ \citep{CHELH_94}.

The fact that the H$\alpha$ profile observed by \citet{PHHSK_90} at day 498
   was not flat topped implies that there was no large cavity free of hydrogen
   at the center of the ejecta, but that hydrogen was mixed deeply down
   to a velocity of $\sim$500\,km\,s$^{-1}$.
\citet{Chu_91} investigated the profiles of hydrogen emission lines at day 350
   and argued that the slowest moving hydrogen was observed at a velocity of
   $\sim$600\,km\,s$^{-1}$.
\citet{KF_98} analyzed the H$\alpha$ profile taken at day 804 and found that
   hydrogen extended into the core to velocities $\le 700$\,km\,s$^{-1}$.
Thus, the presence of hydrogen within the core of heavy elements is confirmed
   by the observational data of SN~1987A.

The hydrodynamic models of SN~1987A based on evolutionary calculations of the
   pre-SN are consistent with the observed light curve when $^{56}$Ni is
   artificially mixed up to velocities of $\sim$4000\,km\,s$^{-1}$
   \citep{SN_90, BLB_00, Utr_04}.
On the other hand, the hydrodynamic simulations of the SN~1987A explosion based
   on nonevolutionary pre-SN models showed in agreement with observations that
   the bulk of $^{56}$Ni is mixed up to velocities of $2500$\,km\,s$^{-1}$
   \citep{Utr_93, Utr_04}.
Photometric data of SN~1987A combined with spectra are the primary source
   of our knowledge of its properties.
Atmosphere models with the steady-state approximation for the ionization
   kinetics produced a too weak H$\alpha$ line compared to the observed line.
After solving the problem of the H$\alpha$ line strength at the photospheric phase
   in the framework of a time-dependent approach \citep{UC_02, UC_05},
   it became possible to simultaneously fit the hydrodynamic and atmosphere models
   to both the photometric and spectroscopic observations with moderate $^{56}$Ni
   mixing up to velocities of $3000$\,km\,s$^{-1}$ \citep{Utr_05}.
To reproduce the broad light curve maximum with hydrodynamic models,
   \citet{Woo_88}, \citet{SN_90}, \citet{BLB_00}, and \citet{Utr_05} mixed
   hydrogen-rich matter down to velocities of about 1700, 800, 1300, and
   600\,km\,s$^{-1}$, respectively.

Besides the direct spectroscopic evidence for moderate $^{56}$Ni mixing
   up to velocities of $3000$\,km\,s$^{-1}$, there are a number of indirect
   indications.
Moderate $^{56}$Ni mixing is supported by the modeling of infrared emission
   lines at late stages from 200 to 700 days.
\citet{LMS_93} showed that the intensity of iron, cobalt, and nickel
   emission lines and their evolution with time are consistent with $^{56}$Ni
   mixing up to velocities of $\sim$2500\,km\,s$^{-1}$, while the same
   observations led \citet{KF_98} to conclude that the iron synthesized during
   the explosion is mixed to velocities of $2000$\,km\,s$^{-1}$.
Using Monte Carlo techniques for calculating the X-ray emission and
   the 847 and 1238\,keV gamma-ray lines, \citet{PW_88} found that a fit to
   the observations required $^{56}$Co mixing up to a velocity of
   3000\,km\,s$^{-1}$.
In addition, Monte Carlo simulations of gamma-ray transport of the 847 and
   1238\,keV lines of $^{56}$Co in the envelope showed that the velocity of
   up to $50\%$ of the total $^{56}$Ni mass should remain below
   1000\,km\,s$^{-1}$ \citep{BR_95}.

\begin{figure*}[t]
\centering
   \includegraphics[width=0.48\hsize, clip, trim=18 153 67 99]{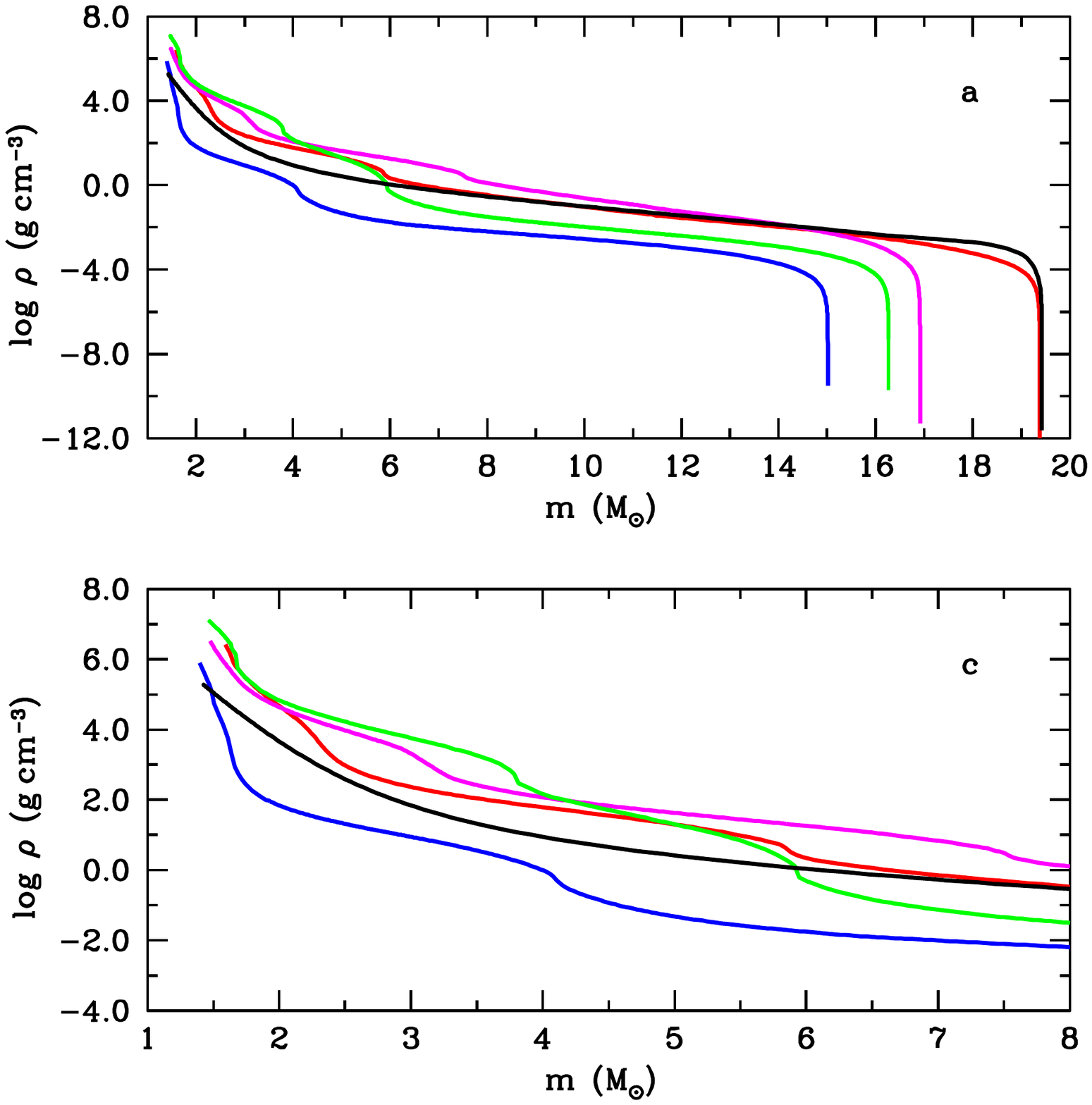}
   \hspace{0.5cm}
   \includegraphics[width=0.48\hsize, clip, trim=18 153 67 99]{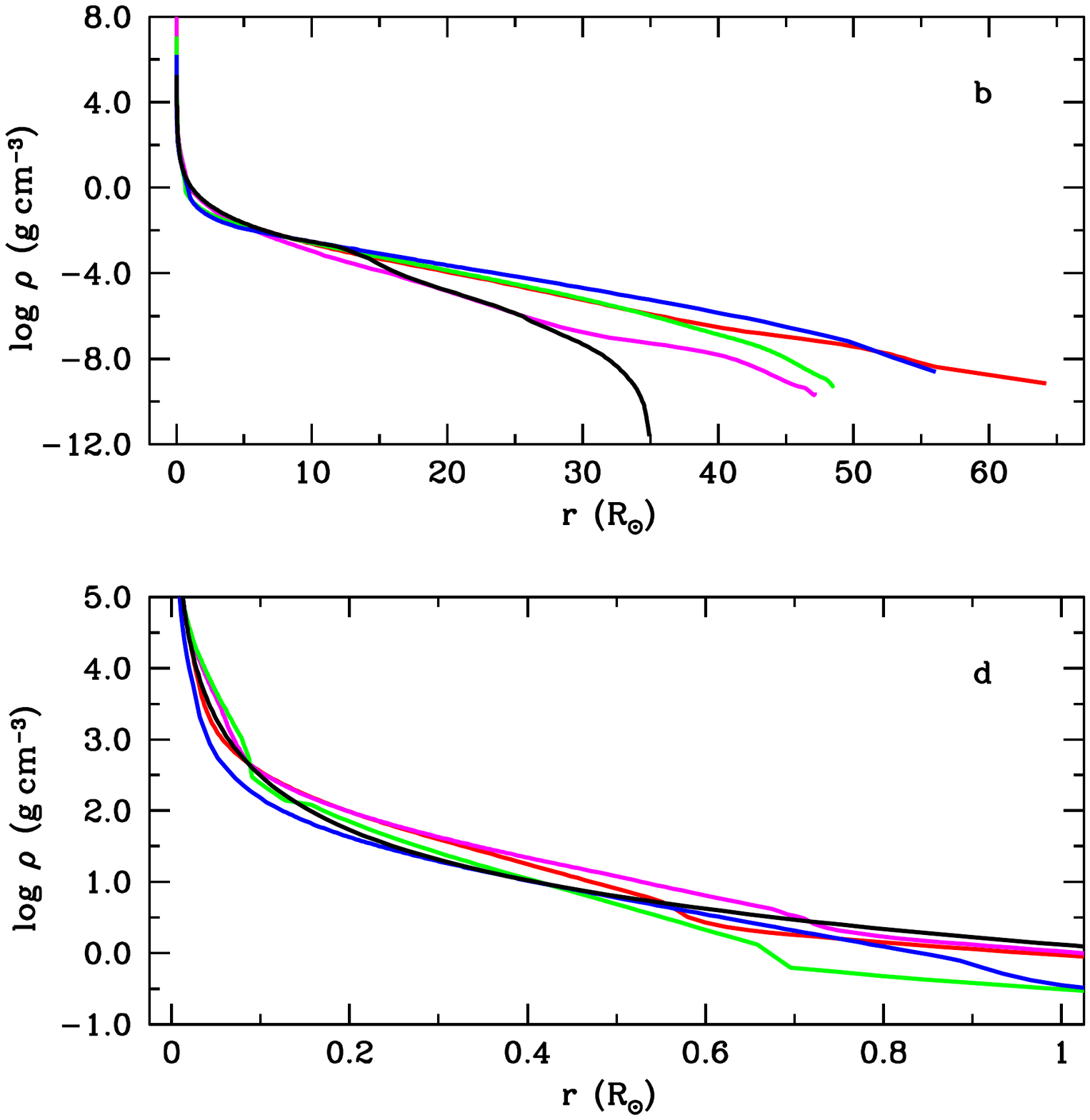}\\
   \caption{%
   Density distributions as functions of interior mass for the whole star
      (Panel \textbf{a}) and the inner region of 8\,$M_{\sun}$
      (Panel \textbf{c}) and as functions of radius for the whole star
      (Panel \textbf{b}) and the inner region of 1\,$R_{\sun}$
      (Panel \textbf{d}) in pre-SN models
      B15 (\emph{blue line\/}), N20 (\emph{green line\/}),
      W18 (\emph{magenta line\/}), and W20 (\emph{red line\/}).
   The central collapsing cores are omitted.
   For a comparison, the \emph{black line} shows the structure of
       a nonevolutionary pre-SN model used by \citet{Utr_05}.
   }
   \label{fig:denmr}
\end{figure*}
These observational data obtained from SN~1987A unambiguously demonstrate
   that the envelope of the pre-SN was substantially fragmented during the
   explosion.
These observations impose serious constraints on the internal problem of
   the SN explosion, and thus stimulated theoretical and numerical work
   on hydrodynamic instabilities, which followed two distinct approaches.
On the one hand, multidimensional modeling of the explosion mechanism was
   attempted \citep{HBC_92, BF_93, MWM_93, HBHFC_94, BHF_95, JM_96, MCB_98},
   mainly with the aim of answering the question as to which extent convective
   instabilities are helpful for generating neutrino-driven explosions.
These simulations were confined to the early shock propagation phase
   up to about one second after core bounce.

On the other hand, hydrodynamic models of the late-time shock propagation and
   the associated formation of Rayleigh-Taylor instabilities in the expanding
   mantle and envelope of the exploding star \citep{AFM_89, FAM_91, MFA_91a,
   MFA_91b, HMNS_90, HMNS_91, HMNS_92, HMNS_94, YS_90, YS_91, HB_91, HB_92,
   HW_94, SIHNS_96, IYN_97, NSS_98, KAR_00} ignored the complications of
   the explosion mechanism. 
Owing to considerable computational difficulties in resolving the relevant
   spatial and temporal scales, all of these investigations relied
   on ad hoc procedures to initiate the explosion.
This usually meant that some simple form of energy deposition (by a piston
   or ``thermal bomb'') was used to generate a shock wave in a pre-SN model.
The subsequent propagation of the blast wave was followed in a one-dimensional
   (1D) simulation until the shock had reached the (C+O)/He or the He/H
   interface.
Then, the 1D model was mapped to a multidimensional grid, and an assumed
   spectrum of seed perturbations was added to the radial velocity field
   to break the spherical symmetry of the problem and to trigger
   the growth of instabilities.
\citet{EYFR_12} somewhat improved the appoach of the studies mentioned above by starting 3D calculations
   shortly after the successful shock revival.
The studies of \citet{YS_90, YS_91}, \citet{NHSY_97, NSS_98} and the
   three-dimensional (3D) calculations of \citet{HFW_03}, who investigated
   the effects of metal mixing on X-ray and gamma-ray spectra formation,
   differ from this approach by making use of parameterized,
   aspherical shock waves.

All of the Rayleigh-Taylor simulations that had been carried out for
   reproducing the mixing in SN~1987A with spherically symmetric initiation
   of the explosion had shared the same problem: with at most
   2000\,km\,s$^{-1}$ their maximum $^{56}$Ni velocities were significantly
   smaller than the observed values of $3000$\,km\,s$^{-1}$ for the bulk of 
   radioactive $^{56}$Ni.
\citet{HB_92} dubbed this problem the ``nickel discrepancy''.
They also speculated that it should disappear when the ``premixing'' of
   the ejecta during the phase of neutrino-driven convection is taken into
   account.
On the other hand, \citet{NSS_98} did not agree with \citeauthor{HB_92}.
They claimed that the nickel discrepancy is resolved if the SN shock is (mildly)
   aspherical.
Similar conclusions were reached by \citet{HFW_03}.
While \citeauthor{NSS_98} as well as \citeauthor{HFW_03} did not rule out
   neutrino-driven convection around the neutron star as an explanation for
   the assumed asphericity of their shocks, \citet{KHO_99}, \citet{WMW_02}, and
   \citet{WWH_02} questioned the neutrino-driven mechanism.
Instead, they speculated about ``jet-driven'' explosions, which might originate
   from magnetohydrodynamic effects in connection with a rapidly rotating
   neutron star.

The controversy described above demonstrated the need for models that link
   observable features at very large radii to the actual energy source and
   the mechanism of the explosion.
\citet{KPJM_00, KPSJM_03, KPSJM_06} provided this link for the standard paradigm
   of neutrino-powered explosions in two-dimensional (2D) geometry.
Their models included a detailed treatment of shock revival by neutrinos,
   the accompanying convection and nucleosynthesis, and the growth of
   Rayleigh-Taylor instabilities at the composition interfaces of the pre-SN
   star after shock passage. 
They employed high spatial resolution by making use of adaptive mesh refinement
   techniques, which is mandatory for highly detailed tracking of the episodes
   of clump formation, mixing, and clump propagation.
\citet{KPSJM_03} showed that the large asymmetries imprinted on the ejecta by
   the violent, nonradial mass motions in the SN core, which precede and 
   accompany the neutrino-driven revival of the blast wave, seed the growth of
   secondary Rayleigh-Taylor instability in the shock-accelerated outer shells
   of the exploding star.
Since the developing Rayleigh-Taylor mushrooms are denser than the surrounding
   gas, they are less decelerated than their environment and can penetrate
   the composition interfaces of the pre-SN star, retaining high velocities as
   the SN ejecta expand.
Thus, they carry freshly synthesized radioactive $^{56}$Ni and other heavy
   elements from the vicinity of the nascent neutron star into the outer stellar
   layers.
\citet{KPSJM_06} demonstrated that massive $^{56}$Ni-dominated clumps can be
   mixed deep into the helium shell and even into the hydrogen layer of the
   disrupted star with terminal velocities of up to $\sim$3000\,km\,s$^{-1}$,
   solving the ``nickel discrepancy'' problem.
In turn, hydrogen can be mixed at the He/H composition interface
   downward in velocity space to only 500\,km\,s$^{-1}$.
Recent 3D hydrodynamic simulations of neutrino-driven
   explosions, for SN~1987A as well, \citep{HJM_10, WJM_10, MJW_12, WJM_13, WMJ_15}
   have confirmed the basic results of the 2D simulations in the
   context of radioactive $^{56}$Ni and hydrogen mixing.
\citet{HJM_10}, in particular, demonstrated that the initial explosion
   asymmetries can be less extreme in 3D than in 2D to explain high $^{56}$Ni
   velocities in the hydrogen layer because drag forces lead to less
   deceleration of 3D metal clumps compared to 2D axisymmetric and, thus,
   toroidal structures.
Recently, \citet{HPO_14} have shown that the convective engine, powering
   the standing accretion shock, is slightly more efficient in their
   3D simulations than in 2D ones.
Requiring the same explosion energy of their SN~1987A model, they needed a 4$\%$
   lower luminosity for their neutrino light bulb in the 3D case than in the 2D case.

In this paper, we revisit the analysis of the peculiar SN~1987A by successively
   combining the capabilities of the internal and external problems.
First, we carry out 3D hydrodynamic simulations of neutrino-driven explosions
   for a set of evolutionary models of the BSG Sanduleak $-69^{\circ}202$
   available to us.
These simulations provide us with a complex morphology of radioactive $^{56}$Ni
   and hydrogen mixing, which retains its global radial features after mapping
   to a spherically symmetric grid.
Second, in spherically symmetric geometry we hydrodynamically model the SN~1987A
   outburst and its light curve.
It is noteworthy that the obtained mixing of radioactive $^{56}$Ni and hydrogen
   is more realistic than the artificial mixing used in all previous
   hydrodynamic studies of SN~1987A.
The present study is a first attempt of a self-consistent modeling of the SN
   development from the revival of the stalled shock to the epoch of
   the radioactive tail of the light curve.

We begin in Sect.~\ref{sec:models} with a brief description of the pre-SN models
   used in our study, the 3D hydrodynamic simulations, and the hydrodynamic
   light curve modeling.
Section~\ref{sec:results} analyzes the results obtained and compares them with
   SN~1987A observations.
In Sect.~\ref{sec:discssn} we discuss our results and their implications for the
   problem of the explosion mechanism, and in Sect.~\ref{sec:conclsns} we
   summarize and conclude.

\section{Model overview}
\label{sec:models}
%
\subsection{Presupernova models}
\label{sec:models-presn}
%
\begin{table}[t]
\caption[]{Presupernova models for blue supergiants.}
\label{tab:presnm}
\centering
\begin{tabular}{@{ } l @{ } c @{ } c @{ } c @{ } c @{ } c @{ } c @{ } c @{ } c @{ } c @{ }}
\hline\hline
\noalign{\smallskip}
 Model & $R_\mathrm{pSN}$ & $M_\mathrm{He}^{\,\mathrm{core}}$ & $M_\mathrm{pSN}$ & $M_\mathrm{ZAMS}$
       & $X_\mathrm{surf}$   & $Y_\mathrm{surf}$   & $Z_\mathrm{surf}$   & $\xi_{1.5}$ & Ref \\
       & $(R_{\sun})$ & $(M_{\sun})$ & $(M_{\sun})$ & $(M_{\sun})$ &
       &              &   $(10^{-2})$      &              &    \\
\noalign{\smallskip}
\hline
\noalign{\smallskip}
 B15   & 56.1 & 4.05 & 15.02 & 15.02      & 0.767 & 0.230 & 0.34 & 0.24 & 1 \\
 N20   & 47.9 & 5.98 & 16.27 & $\sim$20.0 & 0.560 & 0.435 & 0.50 & 0.83 & 2 \\
 W18   & 46.8 & 7.40 & 16.92 & 18.0       & 0.480 & 0.515 & 0.50 & 0.68 & 3 \\
 W20   & 64.2 & 5.79 & 19.38 & 20.10      & 0.738 & 0.256 & 0.56 & 0.78 & 4 \\
\noalign{\smallskip}
\hline
\end{tabular}
\tablefoot{%
The columns give the name of the pre-SN model, its radius, $R_\mathrm{pSN}$;
   the helium-core mass, $M_\mathrm{He}^{\,\mathrm{core}}$;
   the pre-SN mass, $M_\mathrm{pSN}$; the progenitor mass,
   $M_\mathrm{ZAMS}$; the mass fraction of hydrogen, $X_\mathrm{surf}$;
   helium, $Y_\mathrm{surf}$; and heavy elements, $Z_\mathrm{surf}$,
   in the hydrogen-rich envelope at the stage of core collapse;
   the compactness parameter, $\xi_{1.5}$; and the corresponding reference.
}
\tablebib{
(1)~\citet{WPE_88};
(2) \citet{SN_90};
(3) \citet{Woo_07};
(4) \citet{WHWL_97}.
}
\end{table}
We have analyzed a set of pre-SN models, B15, N20, W18, and W20 obtained for
   15, 20, 18, and 20\,$M_{\sun}$ progenitor stars evolved by \citet{WPE_88},
   \citet{SN_90}, \citet{Woo_07}, and \citet{WHWL_97}, respectively
   (Table~\ref{tab:presnm}).
These pre-SN models are used as the initial data in our 3D neutrino-driven
   simulations.
All pre-SN models, except B15, are computed with mass loss. 
In contrast to the other models, model W18 results from the evolution of
   a rotating progenitor.
Mass loss and rotation along with the metal-deficient composition of
   the LMC and the convective mixing favor the formation of BSG-like pre-SN
   models (Table~\ref{tab:presnm}).

\citet{OO_11} studied the spherically symmetric collapse of stellar cores by hydrodynamic
   simulations with a simplified neutrino treatment.
In their study, \citeauthor{OO_11} proposed
   that the fate of collapsing stars, successful explosion with neutron star
   formation or failing SN and black hole formation, can be predicted with
   a single parameter, the compactness of the pre-SN structure at core bounce,
   $\xi_{2.5}$, of the innermost 2.5\,$M_{\sun}$ of a star.
\citet{SW_13} showed that this parameter might be evaluated for the
   ``presupernova model'' defined by the moment when the collapse speed
   reaches 1000\,km\,s$^{-1}$.
They also compared the compactness values for different choices of the enclosed
   mass.
It appears reasonable to assume that the 1.5\,$M_{\sun}$ core of the pre-SN
   is more relevant in connection to the production of radioactive $^{56}$Ni
   and its mixing compared to the value considered by \citeauthor{OO_11}.
For reference, we also give the compactness parameter $\xi_{1.5}$ for
   all of our pre-SN models in Table~\ref{tab:presnm}.

\begin{figure}[t]
   \includegraphics[width=\hsize, clip, trim=20 153 67 215]{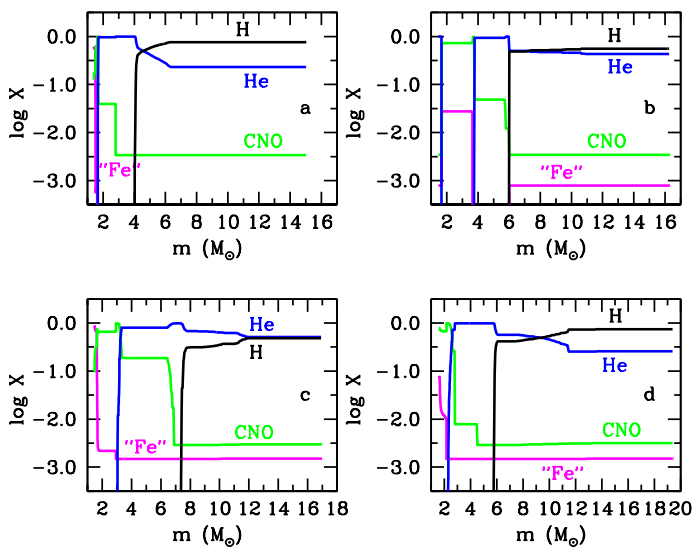}
   \caption{%
   Mass fractions of hydrogen (\emph{black line\/}), helium
      (\emph{blue line\/}), CNO group elements (\emph{green line\/}),
      and iron group elements (\emph{magenta line\/}) in the pre-SN
      models B15 (Panel \textbf{a}), N20 (Panel \textbf{b}),
      W18 (Panel \textbf{c}), and W20 (Panel \textbf{d}).
   }
   \label{fig:chcom}
\end{figure}

\citet{SNK_88} and \citet{Woo_88} showed that the luminosity of the BSG
   Sanduleak $-69^{\circ}202$ is consistent with a star that had a helium core
   of $\approx$6\,$M_{\sun}$ at the time of explosion.
Models N20 and W20 match these pre-explosion observations of SN~1987A, but
   models B15 and W18 have a helium-core mass\footnote
   {We define the helium-core mass as the mass enclosed by the shell
   where the mass fraction of hydrogen $X$ drops below a value of $X=0.01$
   when moving inward from the surface of a star
   (S. E. Woosley, private communication).}
   quite different from the required value (Table~\ref{tab:presnm},
   Figs.~\ref{fig:denmr}a and c).
\citet{KPSJM_06} and \citet{WMJ_15} found that the amount of outward $^{56}$Ni
   mixing and inward hydrogen mixing is sensitive to the structure of
   the helium core and the He/H composition interface.
The different helium cores in the mass range from 4 to 7.5\,$M_{\sun}$
   (Table~\ref{tab:presnm}) allow for a study of this sensitivity for
   BSG progenitors, which also differ in their density distributions versus radius
   (Figs.~\ref{fig:denmr}b and d) and chemical compositions
   (Fig.~\ref{fig:chcom}).

%
\subsection{3D neutrino-driven explosions}
\label{sec:models-3Dexp}
%
Our 3D simulations of supernova explosions are carried out with the explicit
   finite-volume, Eulerian, multifluid hydrodynamics code {\sc Prometheus}
   \citep{FAM_91, MFA_91a, MFA_91b}.
It integrates the multidimensional hydrodynamic equations using dimensional
   splitting \citep{Str_68}, piecewise parabolic reconstruction \citep{CW_84},
   and a Riemann solver for real gases \citep{CG_85}.
Inside grid cells with strong grid-aligned shocks, fluxes computed with the
   Riemann solver are replaced by the AUSM+ fluxes of \citet{Lio_96} 
   to preclude odd-even decoupling \citep{Qui_94}.
The code treats advection of nuclear species by employing the consistent
   multifluid advection scheme of \citet{PM_99}.
 
The code employs an axis-free overlapping ``Yin-Yang'' grid \citep{KS_04}
   in spherical polar coordinates, which was implemented into {\sc Prometheus}
   \citep{WHM_10}.
The Yin-Yang grid relaxes the restrictive CFL-timestep condition and
   avoids numerical artifacts near the polar axis.
Our standard grid configuration consists of
   $400(r)\times47(\theta)\times137(\phi)\times2$ grid cells, corresponding to
   an angular resolution of $2^\circ$ and covering the full $4\pi$ solid angle.
Depending on the evolution phase, hydrostatic equilibrium, a wind-inflow
   condition or an open outflow condition is assumed at the inner grid
   boundary, while a free outflow boundary condition is employed at
   the outer one at all times \citep[see][for details]{WMJ_15}.

We take self-gravity into account by solving Poisson's equation in its
   integral form, using an expansion into spherical harmonics \citep{MS_95}.
The monopole term of the potential is corrected for general relativistic effects
   as described by \citet{SKJM_06} and \citet{AJS_07}.
The inner core of the proto-neutron star (PNS) with densities well above those
   of the neutrinospheric layer is excised and replaced by a point mass at
   the coordinate origin.
The cooling of the excised part of the PNS is then described by neutrino emission
   properties (luminosities and mean spectral energies), which are prescribed as
   time-dependent boundary conditions.
The contraction of the PNS is mimicked by a movement of the inner radial grid
   boundary.
During evolutionary phases significantly later than one second after core
   collapse, the PNS is removed from the simulations and the inner grid
   boundary is placed at successively larger radii \citep[cf.][]{WMJ_15}.

We approximate ``ray-by-ray'' neutrino transport and neutrino-matter interactions
   as in \citet{SKJM_06} by radial integration of the
   1D (spherical), gray transport equation for $\nu_\mathrm{e}$,
   $\bar\nu_\mathrm{e}$, and heavy-lepton neutrinos independently for all
   angular grid directions ($\theta$,\,$\phi$).
This approach allows us to take angular variations of the neutrino
   fluxes produced by the matter accreted onto the newly forming NS into account.
The accretion luminosity adds to the neutrino fluxes imposed spherically and
   symmetrically at the inner grid boundary.
The neutrino spectra are assumed to have Fermi-Dirac shape with chemical
   potentials that are equal to the equilibrium values in optically thick
   regions and constant, limiting values in the free streaming regime.
The spectral temperatures are computed from inverting the ratio of neutrino
   energy and number fluxes, whose transport is solved simultaneously.
Details can be found in \citet{SKJM_06}.
The tabulated equation of state (EoS) of \citet{JM_96} is used to
   describe the stellar fluid.
It includes arbitrarily degenerate and arbitrarily relativistic electrons and
   positrons, photons, and four predefined nuclear species (n, p, $\alpha$, and
   a representative Fe-group nucleus) in nuclear statistical equilibrium.
 
To follow the explosive nucleosynthesis approximately, we solve
   a small $\alpha$-chain reaction network, similar to the network described
   in \citet{KPSJM_03}.
For our simulations we consider 13 $\alpha$-nuclei from $^4$He to $^{56}$Ni and
   an additional ``tracer nucleus''.
The tracer nucleus represents iron-group species that are formed under
   conditions of neutron excess in the neutrino-heated material.
It is produced via the reaction $^{52}$Fe($\alpha$,$\gamma$)$^{56}$Ni within
   grid cells whose electron fraction $Y_\mathrm{e}$ is below 0.49 in our
   simulations.
The tracer thus allows us to keep track of nucleosynthesis in regions with
   neutron excess.
However, it should be noted that some fraction of the tracer material may
   actually be $^{56}$Ni because our approximations in the neutrino transport
   tend to underestimate $Y_\mathrm{e}$ in the neutrino-heated ejecta, while more
   sophisticated energy-dependent neutrino transport can also yield (slightly)
   proton-rich conditions in the neutrino-processed material expelled during
   the early explosion \citep[e.g.,][]{PHWJB_06, FML_06, FWMTL_10, MJM_12}.
The exact $Y_\mathrm{e}$ value of this SN ejecta component, however, remains
   a matter of vivid debate and sensitively depends on uncertain
   aspects of nuclear physics, neutrino reactions, and the dynamics of the
   collapsing stellar core (see, e.g., the review by \citet{Jan_12} and
   new results by \citet{THJ_14}).

Since we are therefore unable to determine the $^{56}$Ni production in our
   models very accurately, we provide maximum and minimum $^{56}$Ni yields,
   depending on whether the tracer material is added to the explosively
   produced $^{56}$Ni mass or not.
The explosively assembled radioactive nickel originates from shock-heated
   ejecta and is computed with our nuclear network.
A ``representative'' mass of radioactive $^{56}$Ni is defined by all
   explosively made nickel plus $50\%$ of the tracer mass
   (cf. Table~\ref{tab:3Dsim}).

The network is solved in grid cells whose temperature is within the range of
   $10^8$\,K -- $8\times10^9$\,K.
We assume that all nuclei are photodisintegrated to $\alpha$-particles at
   temperatures above $8\times10^9$\,K.
This is the high-temperature nuclear statistical equilibrium composition
   compatible with our $\alpha$-network in the absence of free neutrons and
   protons. 
We neglect feedback from the detailed network composition to the EoS and
   the hydrodynamic flow.
This is an acceptable approximation because at SN conditions the contributions
   of nuclei to pressure, energy density, and entropy are dwarfed by those of
   electrons, positrons, and photons.

After the explosions are launched and the explosion energy approaches to near
   its saturation level ($t \sim 1.1-1.3$\,s), we increase the outer boundary
   of the computational grid, which initially resides within the CO core of
   the progenitor stars, to a radius of $r = 10^{14}$\,cm.
Outside of the surface of the progenitors we assume an $r^{-2}$ stellar wind
   density profile.
Enlarging the computational domain allows us to follow the propagation of
   the SN shock wave through the stars and to simulate large-scale mixing
   processes during the subsequent evolution until beyond the SN shock
   breakout from the surface of the progenitors.
In these long-time calculations, we neglect neutrino transport and
   neutrino-matter interactions, and the general-relativistic monopole
   correction to self-gravity.
We switch the EoS table to that of \citet{TS_00}.
We refer to \citet{WMJ_15} for full details of the input physics employed
   in these calculations.
Our 3D long-time simulations of neutrino-driven explosions are stopped
   at $t \sim 44\,000-61\,000$\,s.

The pre-SN models described in Sect.~\ref{sec:models-presn} and
   3D simulations of supernova explosions until at least a time close to shock
   breakout from the star set up the internal problem, which is the explosion
   mechanism itself.

\subsection{Light curves}
\label{sec:models-lcurves}
%
Our radiation hydrodynamics code {\sc Crab} \citep{Utr_04,
   Utr_07} integrates the spherically symmetric equations.
To follow the development beyond our 3D simulations of neutrino-driven
   explosions, we average the 3D hydrodynamic flow and the distribution of
   chemical elements on a spherically symmetric grid at specific times
   and interpolate them onto the computational mass (Lagrangian) grid
   of the 1D simulations.
These data are used as the initial conditions for the external problem of
   the hydrodynamic modeling of the SN outburst.

The SN explosion is normally initiated by a supersonic piston applied to the
   bottom of the stellar envelope at the boundary of the $\sim$1.4\,$M_{\sun}$
   central core, which is removed from the computational mass domain and
   assumed to collapse to become a neutron star.
With the hydrodynamic flow being given by our 3D simulations of neutrino-driven
   explosions, we do not need to mimic the explosion by a supersonic piston at
   the inner boundary and, instead, have to take into account fallback
   onto the young neutron star, i.e., some portion of the matter outside the
   neutron star that initially moves out, but eventually falls back.
To adequately treat fallback and to estimate its mass, we use a ballistic flow
   approximation at the inner boundary neglecting pressure effects
   \citep[e.g.,][]{Che_89}.
This kind of an inner boundary condition describes the infalling matter or the gas
   outflowing but bound to the neutron star.
The innermost zone, to which this boundary condition is applied, is removed
   from the numerical mass grid just when its radius reaches a very small
   value of $0.01\,R_{\sun}$.

The numerical modeling of the SN outbursts employs the implicit, Lagrangian,
   radiation hydrodynamics code {\sc Crab}.
The spherically symmetric hydrodynamic equations with a gravity force and
   radiation transfer equation \citep[e.g.,][]{MM_84} in the one-group
   (gray) approximation are discretized spatially using the method of lines
   \citep[e.g.,][]{HNW_93, HW_96}.
The derived system of ordinary differential equations is stiff and is integrated
   by the implicit method of \citet{Gea_71} with an automatic choice of both
   the time integration step and the order of accuracy of the method.
To automatically capture shocks, the linear and
   nonlinear artificial viscosity of \citet{CSW_98} is added into the
   hydrodynamic equations.
The radiation hydrodynamic equations include additional Compton cooling and
   heating according to \citet{Wey_66}.

The time-dependent radiative transfer equation, written in a comoving frame of
   reference to within terms close to the ratio of the matter velocity
   to the speed of light, is solved as a system of equations for the zeroth
   and first angular moments of the nonequilibrium radiation intensity.
To close this system of moment equations, we use a variable Eddington factor
   that is calculated by directly taking the scattering of
   radiation in the ejecta into account. 
In the inner, optically thick layers of the SN envelope, where thermalization
   of radiation takes place, the diffusion of equilibrium radiation is described
   in the approximation of radiative heat conduction. 
In these layers, local thermodynamic equilibrium (LTE) applies.
The bolometric luminosity of the SN is calculated by including retardation
   and limb-darkening effects.
Taking the latter effects into account is necessary to reproduce
   the bolometric light curve of SN~1987A before and after the broad luminosity
   peak correctly \citep{Utr_04}.

The gamma rays with energies of about 1\,MeV from the decay chain
   $^{56}$Ni $\to ^{56}$Co $\to ^{56}$Fe deposit their energy through
   Compton scattering with free and bound electrons.
The Compton electrons lose their energy through Coulomb heating of free
   electrons, and ionization and excitation of atoms and ions.
The rates of nonthermal heating, excitation, and ionization of atoms
   and ions are taken from \citet{KF_92}.
They are included in the radiation hydrodynamic equations and the equation
   of state.
The gamma-ray transport is calculated with the approximation of an effective
   absorption opacity of 0.06 $Y_\mathrm{e}$\,cm$^2$\,g$^{-1}$.
Positrons are assumed to deposit their energy locally.

Generally, the equation of state for an ideal gas in a nonequilibrium radiation
   field, and for non-thermal excitation and ionization requires solving the
   problem of the level populations and the ionization balance. 
The need of corresponding multiple calculations of the equation of state
   in hydrodynamic modeling forces us to neglect the excited atomic and ionic
   levels, and to restrict our analysis only to the atomic and ionic ground
   states, i.e. to their ionization balance. 
The elements H, He, C, N, O, Ne, Na, Mg, Si, S, Ar, Ca, Fe, and the negative
   hydrogen ion H$^{-}$ are included in the non-LTE ionization balance.
All elements but H are treated with three ionization stages.
The atoms and ions are assumed to consist of the ground state and continuum.
The ionization balance is controlled by the following elementary processes:
   photoionization and radiative recombination, electron ionization and
   three-particle recombination, and non-thermal ionization.
The partition functions are calculated with the polynomial approximation fit
   obtained by \citet{Irw_81}.
The photoionization cross sections of atoms and ions are evaluated with data
   of \citet{VY_95}, and \citet{VFKY_96}.
The electron collisional ionization rates for atoms and ions are computed
   using the approximate formulae of \citet{Vor_97}.
The photoionization cross section data for the negative hydrogen ion are taken
   from \citet{Wis_79}, and the rate coefficient of the electron collisional
   detachment reaction for the negative hydrogen ion from \citet{JLE_87}.

Non-LTE effects are also taken into account when determining the mean
   opacities, the thermal emission coefficient, and the contribution of lines
   to the opacity.
The mean opacities include processes of photoionization, free-free absorption,
   Thomson scattering on free electrons, and Rayleigh scattering on neutral
   hydrogen.
The free-free absorption coefficient is calculated with the effective nuclear
   charge including screening effects \citep{SD_93} and the temperature-averaged
   free-free Gaunt factor from \citet{Sut_98}.
The free-free absorption coefficient of negative hydrogen ions was calculated
   by \citet{BB_87}.
The Rayleigh scattering by hydrogen atoms is calculated using the cross-section
   of \citet{Gav_67} and the exact static dipole polarizability of hydrogen
   from \citet{TP_71}.

\begin{table}[t]
\caption[]{Basic properties of the averaged 3D simulations.}
\label{tab:3Dsim}
\centering
\begin{tabular}{@{ } l @{ } c @{ } c @{ } c @{ } c @{ } c @{ } c @{ } c @{ } c @{ } c @{ }}
\hline\hline
\noalign{\smallskip}
 Model & $M_\mathrm{CC}$ & \phantom{e}$M_\mathrm{env}$ & $E_\mathrm{exp}$
       & $M_\mathrm{Ni}^{\,\mathrm{min}}$ 
       & \phantom{e}$M_\mathrm{Ni}^{\,\mathrm{max}}$
       & \phantom{e}$M_\mathrm{Ni}^{\,\mathrm{rpr}}$
       & \phantom{e}$t_\mathrm{map}^{\,\mathrm{e}}$
       & \phantom{e}$t_\mathrm{map}^{\,\mathrm{l}}$
       & \phantom{e}$t_\mathrm{SB}$ \\
       & \multicolumn{2}{c}{$(M_{\sun})$} & (B)
       & \multicolumn{3}{c}{$(10^{-2}\,M_{\sun})$}
       & \multicolumn{3}{c}{($10^{3}$\,s)} \\ 
\noalign{\smallskip}
\hline
\noalign{\smallskip}
B15-1 & 1.26 & 14.19 & 1.15 & 2.87 &  8.25 & 5.58 & 6.62 & 61.2 & 7.81 \\
B15-2 & 1.25 & 14.21 & 1.40 & 3.11 &  9.36 & 6.23 & 4.83 & 61.2 & 7.10 \\
B15-3 & 1.25 & 14.29 & 2.59 & 4.81 & 11.08 & 7.94 & 3.99 & 44.6 & 5.29 \\
N20-P & 1.46 & 14.72 & 1.67 & 4.16 & 12.11 & 8.13 & 3.77 & 56.9 & 5.36 \\
N20-C & 1.46 & 14.72 & 1.67 & 4.16 & 12.11 & 8.13 & 3.71 & 56.9 & 5.35 \\
W18   & 1.40 & 15.52 & 1.36 & 3.67 & 12.97 & 8.32 & 3.40 & 55.2 & 4.26 \\
W20   & 1.50 & 17.92 & 1.45 & 4.10 & 13.04 & 8.57 & 4.21 & 61.2 & 6.67 \\
\noalign{\smallskip}
\hline
\end{tabular}
\tablefoot{%
The computed models are based on the corresponding pre-SN models
   of Table~\ref{tab:presnm}, except for model N20-C, which has the same
   composition of metals as model N20-P with its hydrogen mass fraction
   enhanced to $X_\mathrm{surf}=0.735$ at the expense of helium in
   the hydrogen-rich envelope.
$M_\mathrm{CC}$ is the mass of the collapsing core;
   $M_\mathrm{env}$, the ejecta mass;
   $E_\mathrm{exp}$, the explosion energy;
   $M_\mathrm{Ni}^{\,\mathrm{min}}$, the mass of radioactive $^{56}$Ni produced
   directly in shock-heated ejecta by our reaction network;
   $M_\mathrm{Ni}^{\,\mathrm{max}}$, the aggregate mass of directly produced
   $^{56}$Ni and tracer nucleus; and
   $M_\mathrm{Ni}^{\,\mathrm{rpr}}$, the representative radioactive $^{56}$Ni
   mass containing the whole directly produced $^{56}$Ni and half of the
   tracer nucleus.
$t_\mathrm{map}^{\,\mathrm{e}}$ and $t_\mathrm{map}^{\,\mathrm{l}}$
   are the two moments at which the 3D simulations are mapped to a spherically
   symmetric grid.
$t_\mathrm{SB}$ is the epoch of shock breakout in the early-time 1D
   simulations.
}
\end{table}
The contribution of spectral lines to the opacity in an expanding medium with 
   a velocity gradient is evaluated by the generalized formula of \citet{CAK_75}
   and is treated as pure scattering.
Oscillator strengths of lines are taken from the line database of \citet{Kur_02}
   containing nearly 530\,000 lines.
Energy level data are from the atomic spectra database of the National
   Institute of Standards and Technology.
Atomic and ionic level populations are determined by the Boltzmann formulae and
   the Saha equations for a mixture of all elements from H to Zn with the local
   nonequilibrium radiation temperature.

\begin{figure*}
\centering
   \includegraphics[width=0.245\hsize, clip, trim=80 80 80 80]{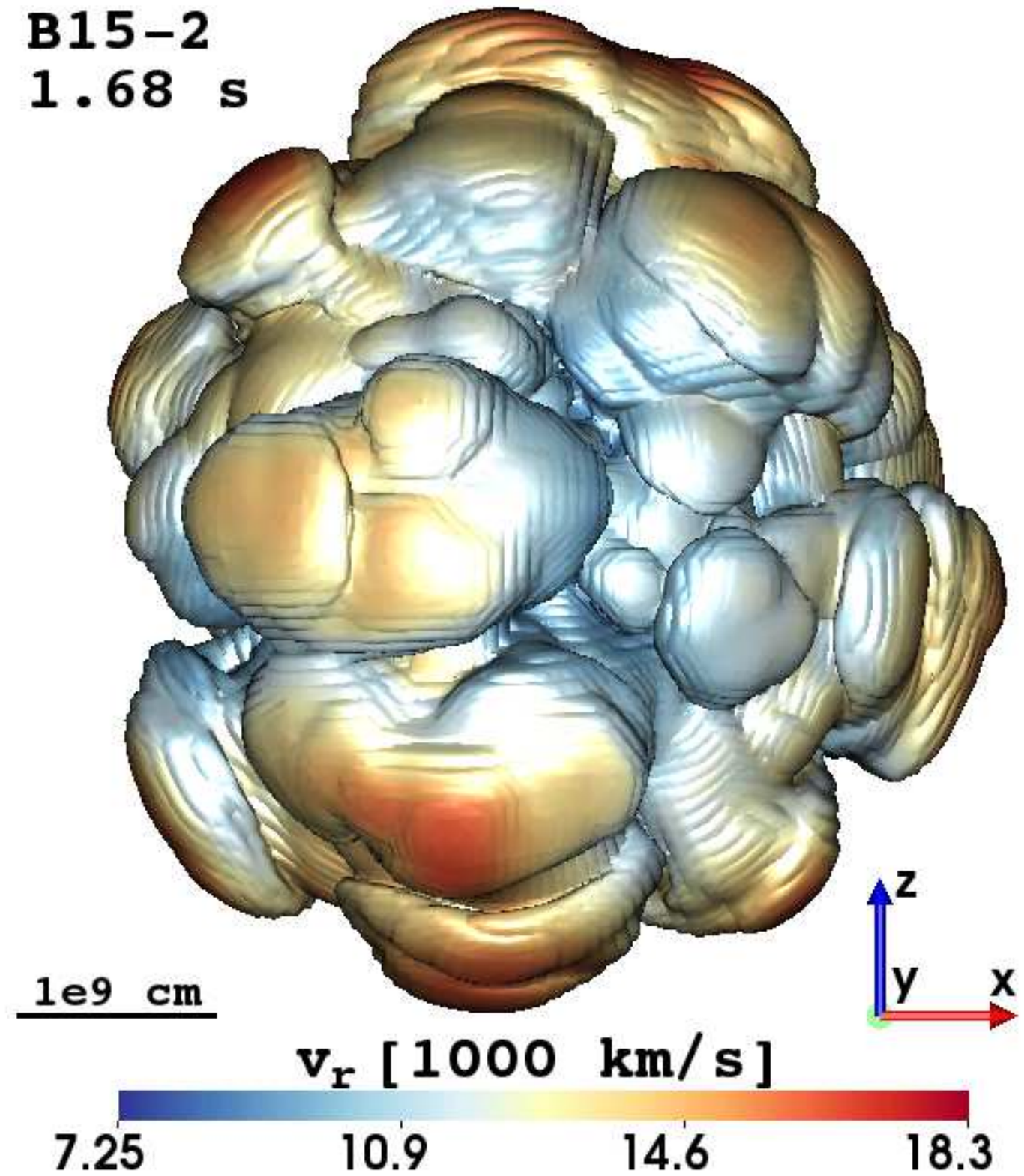}
   \includegraphics[width=0.245\hsize, clip, trim=80 80 80 80]{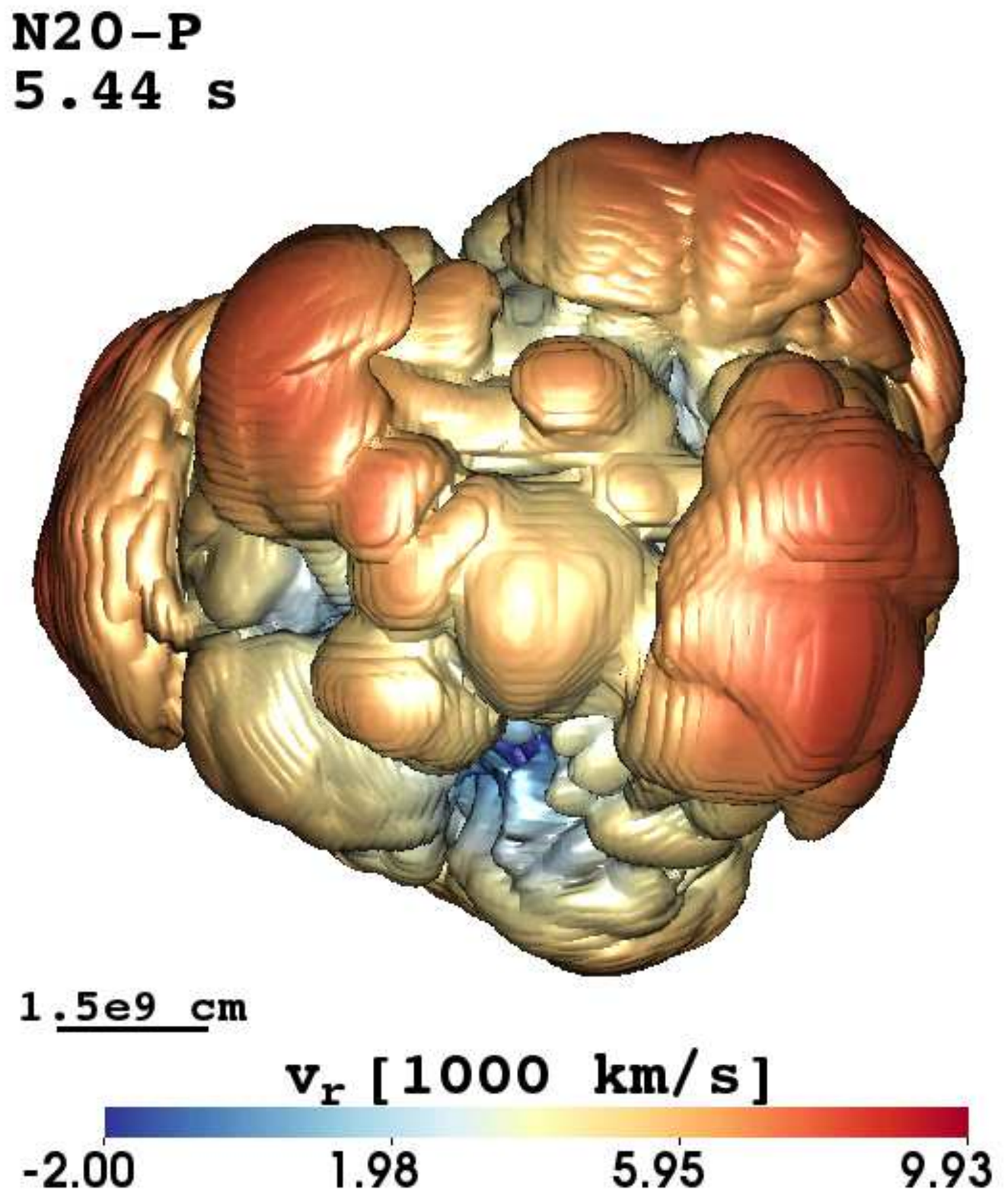}
   \includegraphics[width=0.245\hsize, clip, trim=80 80 80 80]{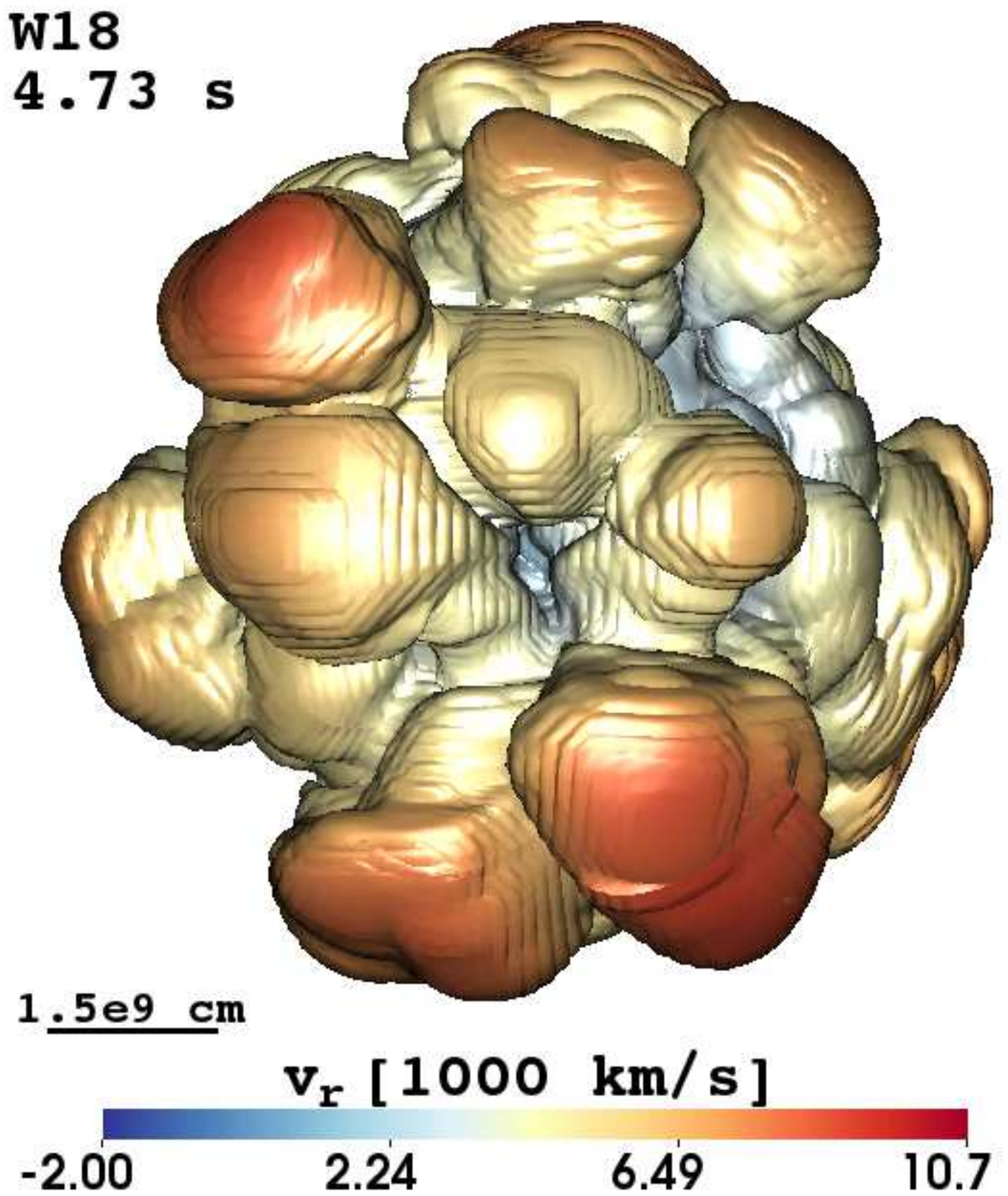}
   \includegraphics[width=0.245\hsize, clip, trim=80 80 80 80]{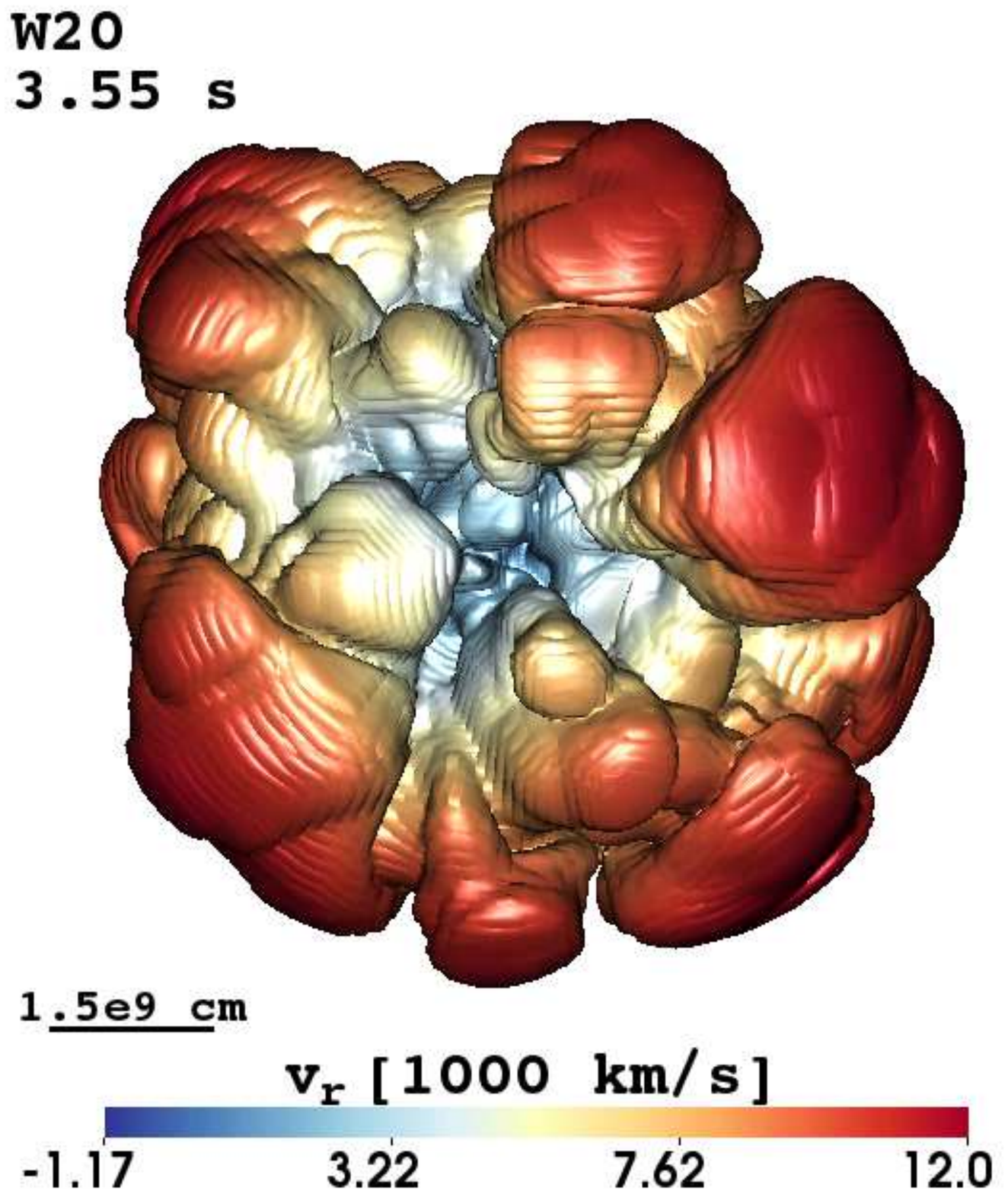}\\
   \hspace{0.0cm}
   \includegraphics[width=0.245\hsize, clip, trim=80 80 80 80]{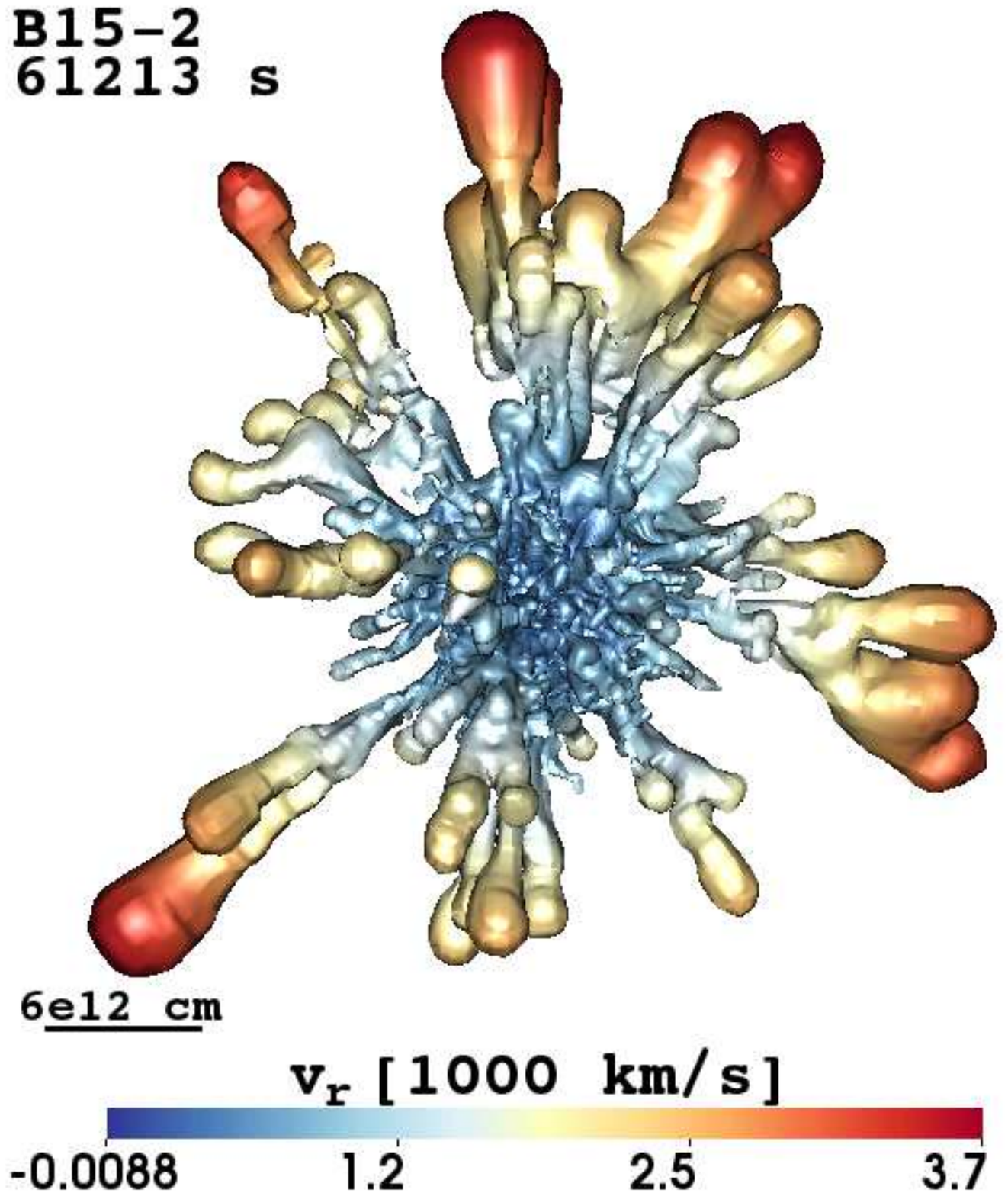}
   \includegraphics[width=0.245\hsize, clip, trim=80 80 80 80]{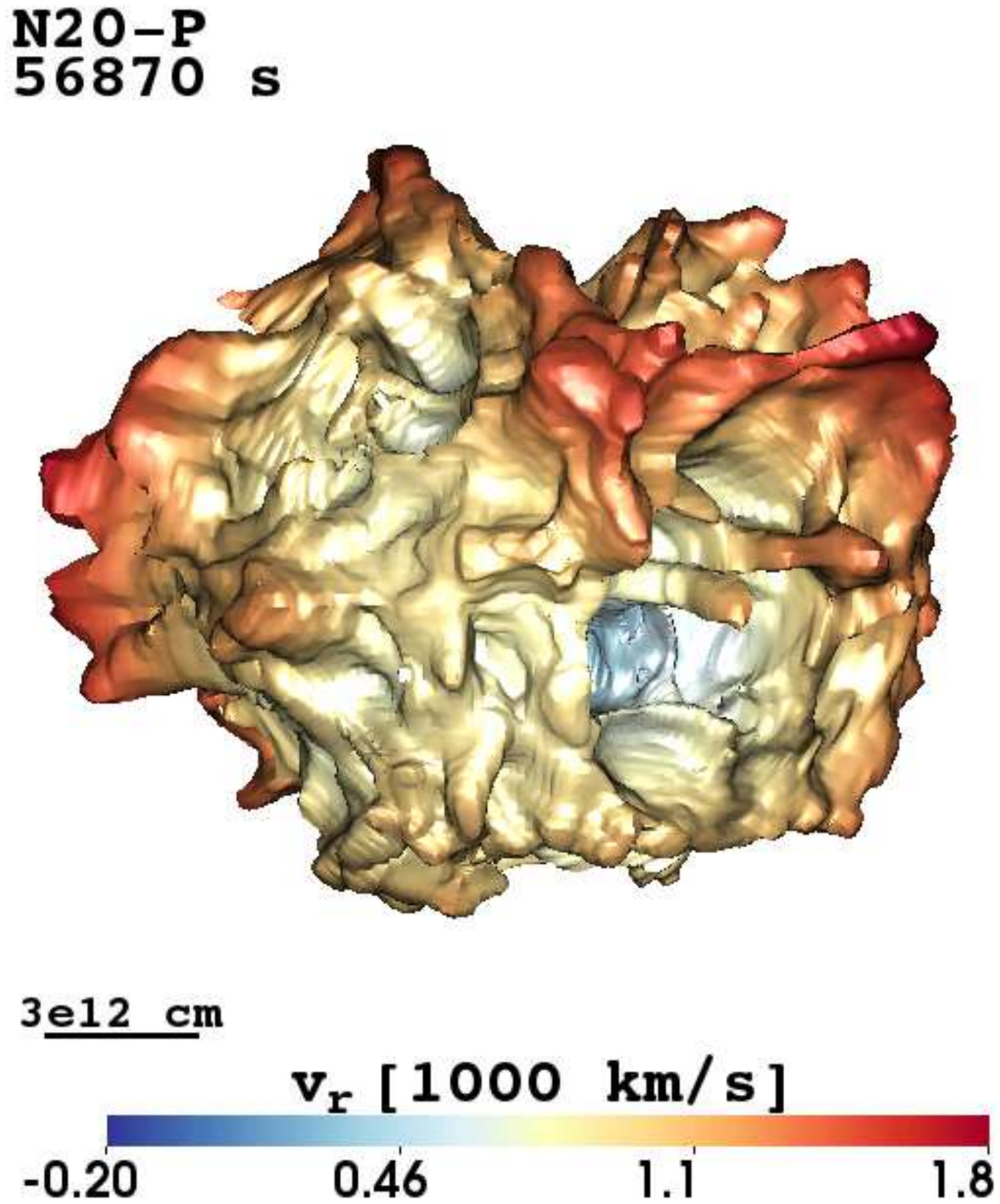}
   \includegraphics[width=0.245\hsize, clip, trim=80 80 80 80]{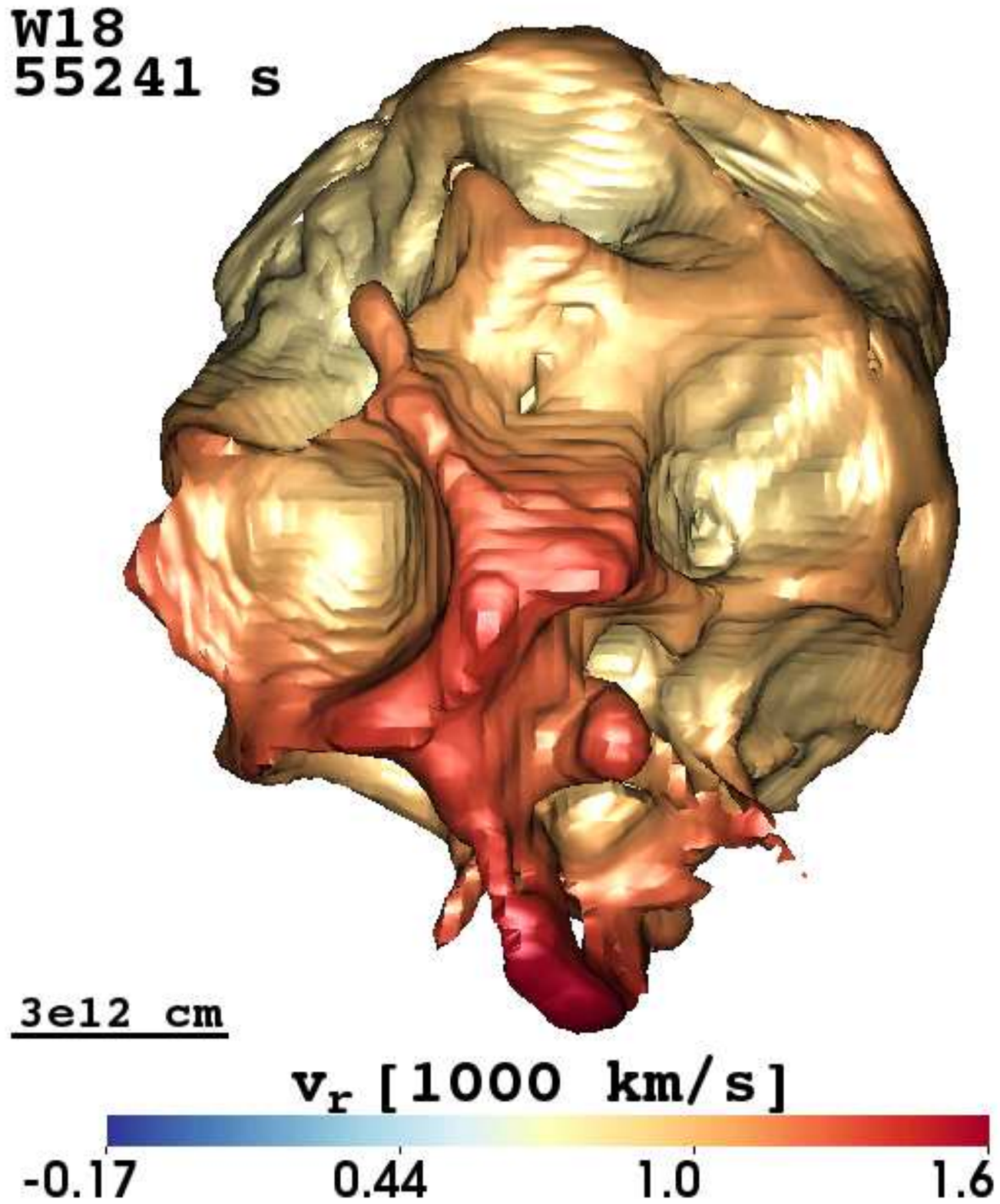}
   \includegraphics[width=0.245\hsize, clip, trim=80 80 80 80]{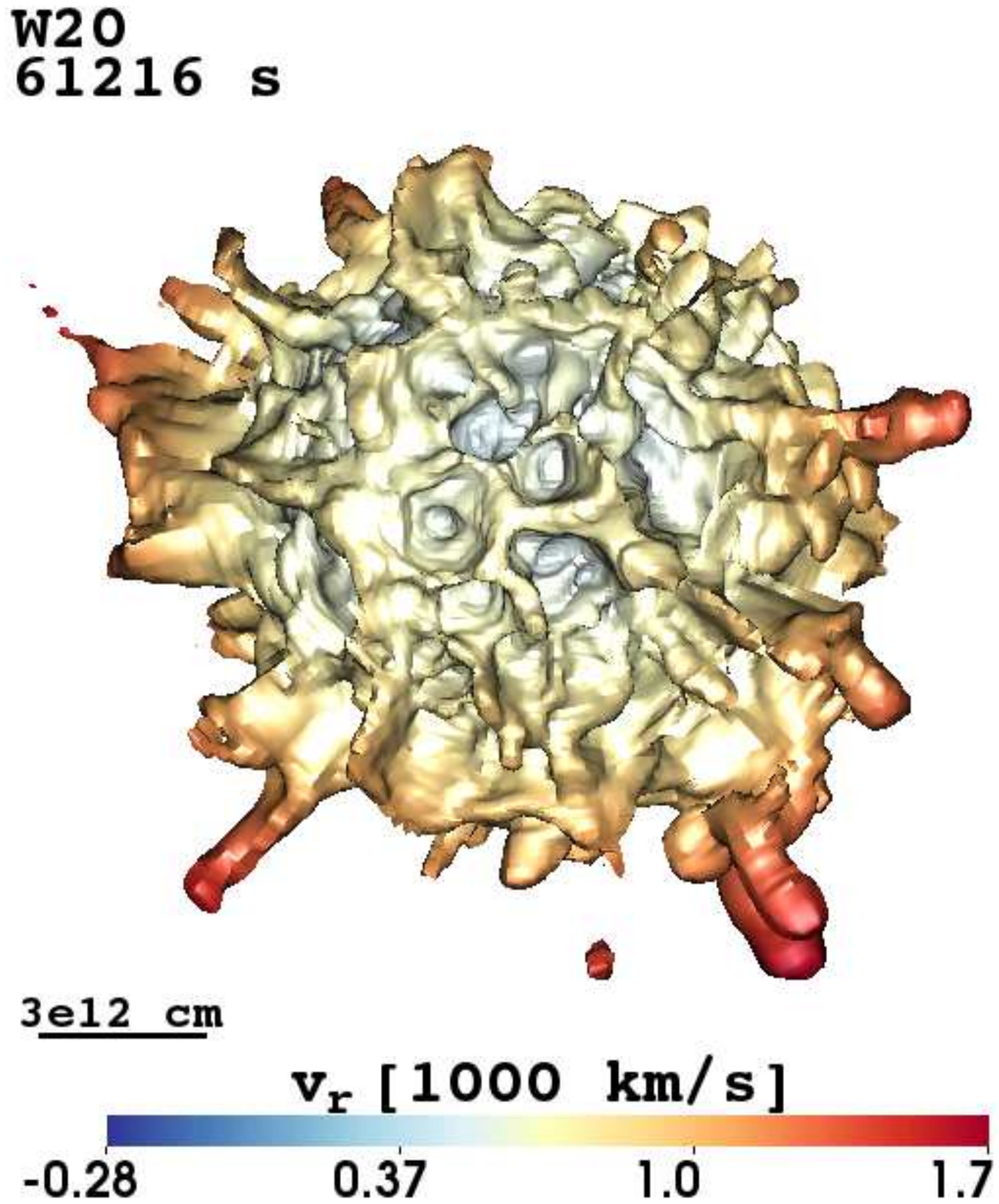}\\
   \caption{%
   Morphology of radioactive $^{56}$Ni-rich matter produced by explosive
      burning in shock-heated ejecta.
   The snapshots display isosurfaces where the mass fraction of $^{56}$Ni plus
      the neutron-rich tracer nucleus equals $3\%$.
   The isosurfaces are shown for 3D models B15-2, N20-P,
      W18, and W20 at two different epochs: shortly before the SN
      shock crosses the C+O/He composition interface in the progenitor star
      at $t=1.68, 5.44, 4.73$, and $3.55$\,s after bounce (upper row) and
      long after the shock breakout at $t=61\,213, 56\,870, 55\,241$, and
      $61\,216$\,s (lower row), respectively.
   The colors give the radial velocity on the isosurface, the color coding
      being defined at the bottom of each panel.
   In the top left corner of each panel, we give the name of the model and
      the post-bounce time of the snapshot.
   The size of the displayed volume and the clumps can be estimated from
      the yardsticks given in the lower left corner of each panel.
   One notices a striking difference between model B15-2 and the other models in
      the final morphology of the $^{56}$Ni-rich ejecta, which arises from
      their specific progenitor structures and the influence of the latter
      on the unsteady SN shock propagation.
   }
   \label{fig:3D_models}
\end{figure*}
The main shortcomings of our radiation hydrodynamics code {\sc Crab}
   are the assumption of spherical symmetry and the gray approximation
   for the photon radiation transport.
The latter approximation is, however, sufficiently accurate to reproduce
   the bolometric light curve of SN~1987A.
Other codes used in the modeling of supernova light curves are, for
   example, {\sc PHOENIX} \citep[see e.g.][]{HB_14} and {\sc SEDONA}
   \citep{KTN_06}, both of which are multidimensional, time-dependent,
   non-LTE, multi-wavelength radiation transport codes, but no radiation
   hydrodynamics codes.

\section{Results}
\label{sec:results}
%
A set of 3D neutrino-driven explosion simulations with our four pre-SN models
   B15, N20, W18, and W20 (Table~\ref{tab:presnm}) as initial data is carried out.
Basic properties of the averaged 3D simulations for seven computed models are
   listed in Table~\ref{tab:3Dsim}.
We map our 3D simulations to a spherically symmetric grid at two different epochs:
   at an early time $t_\mathrm{map}^{\,\mathrm{e}}$, prior to shock breakout,
   and at a late time $t_\mathrm{map}^{\,\mathrm{l}}$, long after shock breakout.
We use these 1D models mapped from the 3D data as the initial conditions
   to simulate the SN outburst with the spherically symmetric code {\sc Crab}.
From this point on we will refer to these 1D hydrodynamic models as ``early-time or
   late-time 1D simulations (models)'' as a continuation of the corresponding
   3D simulations.
We define the explosion energy $E_\mathrm{exp}$ as the sum of the total (i.e.,
   internal plus kinetic plus gravitational) energy of all grid cells
   at the mapping moment.
Throughout this paper, we employ the energy unit
   $1\,\mathrm{bethe}=1\,\mathrm{B}=10^{51}$\,erg.

\subsection{3D simulations of the first day of explosion}
\label{sec:results-3dtosph}
%
For completeness we outline the development of neutrino-driven
    explosions after core bounce \citep[see, e.g.,][for details]{KPSJM_03,
    KPSJM_06}.
As an illustrative example, we consider our reference model B15-2
   \citep[see][for detailed results and discussion]{WMJ_15}.
Neutrino heating around the neutron star triggers violent buoyancy and mass
   overturn, first visible at about $t=65$\,ms after bounce when Rayleigh-Taylor
   mushrooms begin to grow from the imposed seed perturbations in the
   convectively unstable layer that builds up within the neutrino-heating
   region between the gain radius and the stalled SN shock.
These high-entropy bubbles start rising, growing, merging, partially collapsing
   again, and emerge once more to inflate to larger sizes.
Supported by convective overturn and global shock motions, the delayed,
   neutrino-driven explosion sets in at roughly $t=164$\,ms after bounce.

The value of the explosion energy is determined by the isotropic neutrino
   luminosity at the inner boundary and its prescribed temporal evolution
   plus the accretion luminosity, which is regulated by the mass accretion
   rate of the considered collapsing stellar core and by the gravitational
   potential of the accreting neutron star, which depends on the contraction
   of the inner boundary.

Once the shock wave has been launched by neutrino heating, the further
   evolution of the explosion depends strongly on the density profile of
   the pre-SN.
It is known that the shock decelerates whenever it encounters a density profile
   that falls off with increasing value $\rho r^3$, while it accelerates for density
   profiles with decreasing value $\rho r^3$.
Since the density structure of pre-SN models cannot be described by a single
   power law (Figs.~\ref{fig:denmr}b and d), the shock propagates
   nonmonotonically.
At the locations of the Si/O, (C+O)/He, and He/H interfaces
   (Fig.~\ref{fig:chcom}), the radial dependence of the value of $\rho r^3$
   varies, i.e., the velocity of the shock increases when the shock approaches
   a composition interface and decreases after the shock has crossed the
   interface.

The shock wave first reaches the Si/O interface at around $t=252$\,ms after
   bounce.
When it crosses the (C+O)/He interface at $t=1.68$\,s after
   bounce, the maximum speed of the $^{56}$Ni-rich matter is as high as
   $\approx$18\,300\,km\,s$^{-1}$ (Fig.~\ref{fig:3D_models}, left panel
   in upper row).
Thereafter, a rapid decrease of the shock velocity occurs when the shock enters
   the He-layer, and a reverse shock forms at around $t=15$\,s after core
   bounce.
Once the main shock has passed the He/H interface around $t=66$\,s after bounce,
   the evolution resembles that after the crossing of the (C+O)/He interface
   and another reverse shock forms below the He/H interface at around
   $t=700$\,s.
A few of the fastest Rayleigh-Taylor plumes have actually reached the He/H
   interface before this strong reverse shock has formed, and thereby they have
   escaped an interaction with this shock in stark contrast to the situation in
   the other models N20-P, W18, and W20 (Fig.~\ref{fig:3D_models}).
The motion of the fastest metal clumps relative to the background flow never
   becomes supersonic, and the deceleration enhanced by supersonic drag is absent.
Hence, the maximum velocities of iron-group nuclei and other elements decrease
   only slightly at late times, and the fastest $^{56}$Ni-rich clumps move
   with velocities of $\approx$3700\,km\,s$^{-1}$ at $t=61\,213$\,s after bounce
   (Fig.~\ref{fig:3D_models}, first panel in lower row).
In contrast, the other mushrooms penetrate through the reverse shock and move
   supersonically relative to the ambient medium.
As a result they dissipate a large fraction of their kinetic energy
   in bow shocks and strong acoustic waves, and form an almost spherical
   distribution with velocities up to $\approx$1000\,km\,s$^{-1}$ by
   $t=61\,213$\,s after bounce.

An unsteady shock propagation through the pre-SN gives rise to
   Rayleigh-Taylor unstable (opposite) pressure and density gradients
   at the composition interfaces.
A global deformation of the main shock enhances strong inward
   mixing of hydrogen: the aspherical shock deposits large amounts of vorticity
   into the He/H interface layer and triggers the growth of a strong
   Richtmyer-Meshkov instability.
The latter creates huge hydrogen pockets at the He/H interface intensifying
   mixing of hydrogen-rich matter down to the center.
It should be emphasized that this instability plays a less important role
   in the 3D simulations of \citet{HJM_10} and \citet{WMJ_15} than in
   the 2D models of \citet{KPSJM_03,KPSJM_06}.
The 3D dynamics of outward mixing of radioactive $^{56}$Ni and inward mixing
   of hydrogen-rich matter in the SN models was studied in detail by 
   \citet{WMJ_15}.

\begin{figure}[t]
   \includegraphics[width=\hsize, clip, trim=47 153 47 195]{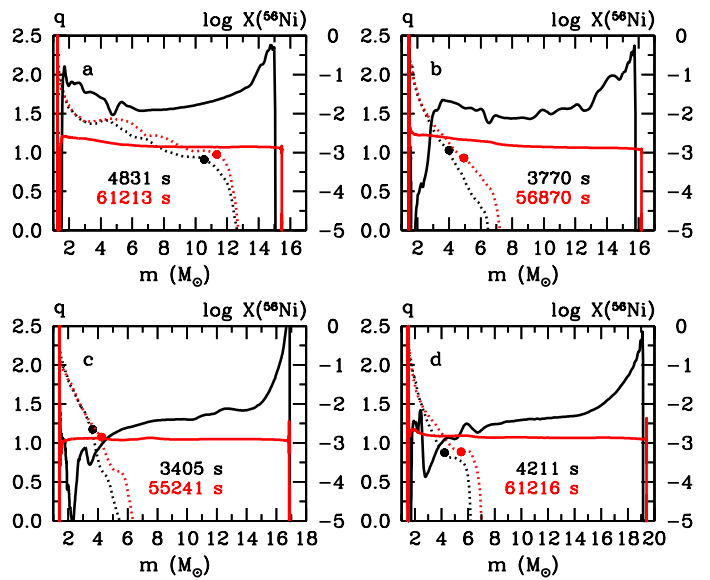}
   \caption{%
   The approach to homologous expansion is shown by comparing
   the early-time (\emph{black lines\/}) and late-time
      (\emph{red lines\/}) 3D simulations of models
      B15-2 (Panel \textbf{a}), N20-P (Panel \textbf{b}),
      W18 (Panel \textbf{c}), and W20 (Panel \textbf{d}).
   The \emph{solid lines} give the homology index  
      $q={\partial \ln v}/{\partial \ln r}$,
      the \emph{dotted lines} the $^{56}$Ni mass fraction.
   \emph{Bullets} mark the outer boundary of the bulk of $^{56}$Ni
      containing $99\%$ of the total $^{56}$Ni mass.
   }
   \label{fig:aphexp}
\end{figure}
During the propagation of the main shock before its breakout, the shock
   deformation is imprinted on the ejecta and determines the morphology of
   the outer layers.
To measure the asphericity of the outer layers in the late-time
   3D simulations of models B15-2, N20-P, W18, and W20, we approximated
   the density isosurface with an ellipsoid in the vicinity of the photosphere
   position found for the averaged 3D simulations.
The maximum ratios of the semiaxes for the approximations thus obtained are
   1.055, 1.058, 1.030, and 1.025, respectively.

\subsection{Approach to homologous expansion}
\label{sec:results-ahexp}
%
As pointed out above, there is observational evidence for macroscopic mixing
   occurring during the explosion of SN~1987A.
Mixing of radioactive $^{56}$Ni out of the center and hydrogen-rich matter
   down to the center is required in order to reproduce both the smooth rising
   part of the bolometric light curve and the width of the broad maximum.
Because macroscopic mixing continues until a phase when the ejecta reach
   a homologous expansion, we are interested in carrying out our 3D simulations
   of SN explosions as long as possible.
Homologous expansion occurs when the contribution of pressure and gravity to
   the momentum equation may be neglected.
A convenient measure to control the approach of the hydrodynamic flow to
   homologous expansion is the effective homology index
   $q={\partial \ln v}/{\partial \ln r}$, which has to go to unity.

We estimate the proximity to homologous expansion by comparing the averaged
   3D simulations at the two different mapping epochs (Table~\ref{tab:3Dsim},
   Fig.~\ref{fig:aphexp}).
At the early-time mapping epoch, $t_\mathrm{map}^{\,\mathrm{e}}$, well prior
   to the phase of shock breakout, the hydrodynamic flow is far from
   homologous expansion in all four models B15-2, N20-P, W18, and W20 because
   the index $q$ varies over a wide range around unity.
It implies that outward mixing of radioactive $^{56}$Ni and inward mixing
   of hydrogen-rich matter in the ejecta will continue until complete homology
   is reached.
By the late-time mapping epoch, $t_\mathrm{map}^{\,\mathrm{l}}$, the index $q$
   is very close to unity in all models, and the ejecta in the 3D
   neutrino-driven simulations of these models are almost in homologous
   expansion.
It is clear that between $t_\mathrm{map}^{\,\mathrm{e}}$ and
   $t_\mathrm{map}^{\,\mathrm{l}}$ macroscopic mixing of radioactive $^{56}$Ni
   will continue (Fig.~\ref{fig:aphexp}).
For this reason, the 3D neutrino-driven simulations mapped at the early-time
   epoch are less reliable than those evolved to the late-time mapping epoch.
At $t_\mathrm{map}^{\,\mathrm{l}}$ the index $q$ is closer to unity in the outer
   layers of the ejecta than in the inner layers because the contribution of
   pressure and gravity to the momentum equation is smaller in the outer layers
   than in the inner layers.
At $t_\mathrm{map}^{\,\mathrm{l}}$ the bulk of $^{56}$Ni containing $99\%$ of
   the total $^{56}$Ni mass expands almost homologously in all four models
   (Fig.~\ref{fig:aphexp}).

However, to predict the subsequent evolution of the nickel ejecta, one
   would have to take additional mixing into account that might result
   from local radioactive decay heating (see also Sect.~\ref{sec:discssn}).
\citet{HB_91} performed 2D simulations of hydrodynamic instabilities and mixing
   in SN~1987A during the period from 400 min to 90 days.
They showed that $^{56}$Ni and subsequent $^{56}$Co decay leads to a relatively
   modest boost of the peak radial velocities of these elements by
   a few hundred km\,s$^{-1}$.
Because of this relatively small change in the velocities of iron-group
   elements by radioactive decay heating, we consider our 3D models
   mapped at $t_\mathrm{map}^{\,\mathrm{l}}$ as a good proxy of the final state
   with respect to mixing and velocity distributions of heavy elements.
After mapping the peak radial velocity of the $^{56}$Ni-rich 
   matter increases by radioactive decay heating, which is treated 
   adequately in the {\sc Crab} code.

\begin{figure}[t]
   \includegraphics[width=\hsize, clip, trim=35 153 40 243]{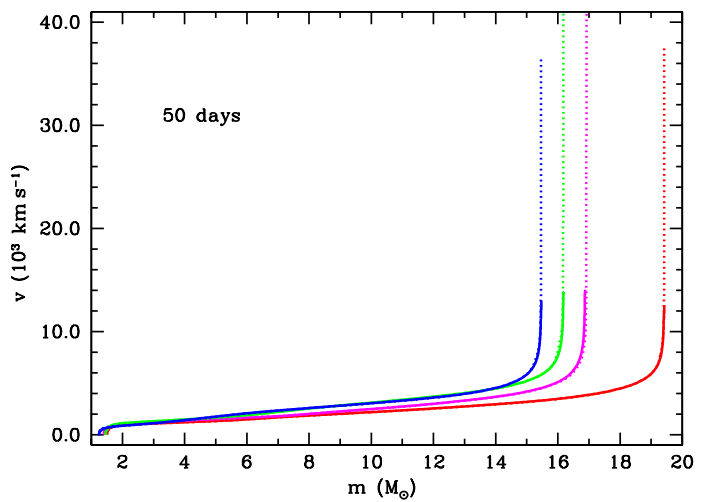}
   \caption{%
   Velocity of the ejecta as function of interior mass
      in models B15-2 (\emph{blue line\/}), N20-P (\emph{green line\/}),
      W18 (\emph{magenta line\/}), and W20 (\emph{red line\/})
      at $t=50$ days for the early-time (\emph{dotted lines\/}) and
      late-time (\emph{solid lines\/}) 1D simulations.
   }
   \label{fig:velm}
\end{figure}
We first analyze the hydrodynamic flow, which is characterized, among other
   factors, by the velocity profile.
In the homologous expansion phase, the expansion velocity of the outermost
   layers in models B15-2, N20-P, W18, and W20 exceeds a value of
   $\approx$36\,000\,km\,s$^{-1}$ for the early-time 1D simulations
   (Fig.~\ref{fig:velm}).
This is consistent with observations of the radio remnant of SN~1987A over
   the period from 1987 to 1990--1991, implying a mean expansion speed of
   $\approx$38\,000\,km\,s$^{-1}$ \citep{GSM_07}.
At the same time, the late-time 1D models result in
   an expansion velocity of the outermost layers that is a factor of three lower
   than that for the early-time 1D models (Fig.~\ref{fig:velm}).

\begin{figure}[t]
   \includegraphics[width=\hsize, clip, trim=18 153 67 210]{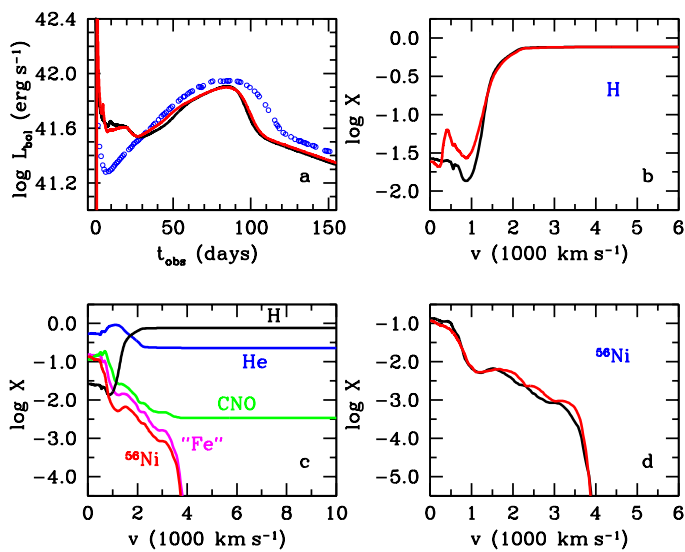}
   \caption{%
   Comparison of the early- and late-time 1D simulations for model B15-2
      with the representative radioactive $^{56}$Ni mass (Table~\ref{tab:3Dsim}).
   Panel \textbf{a}: bolometric light curves of the early-time
      (\emph{black line\/}) and late-time (\emph{red line\/}) 1D simulations
      compared with the observations of SN~1987A obtained by
      \citet{CMM_87, CWF_88} (\emph{open circles\/}).
   Panels \textbf{b} and \textbf{d}: mass fractions of hydrogen and
      radioactive $^{56}$Ni, respectively, as functions of velocity for
      the early-time (\emph{black line\/}) and late-time (\emph{red line\/})
      1D simulations.
   Panel \textbf{c}: mass fractions as functions of velocity for
      the early-time 1D simulation.
   The \emph{red line\/} is the mass fraction of radioactive $^{56}$Ni
      (see also Fig.~\ref{fig:chcom}).
   All profiles of the mass fractions are given at $t=50$ days. 
   }
   \label{fig:elts}
\end{figure}

This disagreement between the early- and late-time runs is a numerical artifact
   that originates from the interaction of the shock with an assumed
   stellar wind, surrounding the pre-SN model.
For the Eulerian simulations with the {\sc Prometheus} code, an ambient medium
   (e.g., in the form of a stellar wind) is necessary to follow the 3D
   neutrino-driven simulations after the shock breakout, while this assumption
   is not needed for the Lagrangian simulations with the {\sc Crab} code.
It turns out that the chosen stellar wind is too dense to reproduce
   the shock velocities after breakout correctly.
Unfortunately, because of the computational cost of 3D simulations, we are not
   able to iterate the stellar wind parameters to minimize the
   influence of the wind on the velocity of the outermost layers after shock
   breakout.
Generally speaking, there is no reason to make efforts to eliminate this
   artifact because even in the case of the early-time 1D runs the light
   curve is inconsistent with the observations of SN~1987A during the first
   30 days (see, for example, for model B15-2 in Fig.~\ref{fig:elts}a).

We now address the mixing of radioactive $^{56}$Ni and hydrogen during the
   explosion and the light curve properties for the early- and late-time
   1D simulations of model B15-2.
Figure~\ref{fig:elts}c shows the mass fractions of hydrogen, helium,
   CNO group elements, and iron group elements as functions of velocity for
   the early-time 1D simulations.
Between $t_\mathrm{map}^{\,\mathrm{e}}$ and $t_\mathrm{map}^{\,\mathrm{l}}$,
   macroscopic mixing in the
   ejecta increases both the mass fraction of hydrogen in the metal core
   (Fig.~\ref{fig:elts}b) and that of radioactive $^{56}$Ni in the outer part
   of the mixing zone, with the maximum $^{56}$Ni velocity remaining unchanged
   (Fig.~\ref{fig:elts}d).
These changes may be considered as insignificant compared to the dramatic
   evolution of the hydrodynamic flow during that period (Fig.~\ref{fig:aphexp}). 
In addition, this result justifies a terminal time for 3D simulations of
   the order of $55\,000-60\,000$\,s after bounce.

A redistribution of radioactive $^{56}$Ni to higher velocities in the mixing zone
   slightly increases the bolometric luminosity relative to the early-time 1D
   run between $\sim$30 days and $\sim$60 days (Fig.~\ref{fig:elts}a), at which
   time the radioactive decay of $^{56}$Ni and $^{56}$Co nuclides becomes
   dominant in powering the luminosity.
An increase of the mass fraction of hydrogen in the metal core enlarges the
   opacity and the optical depth of the inner part of the ejecta and,
   consequently, the photon diffusion time, making the dome-like maximum of
   the light curve slightly lower and wider (Fig.~\ref{fig:elts}a).
Both effects tend to improve the match between computed and
   observed light curves, and therefore we focus mainly on hydrodynamic models for
   the late-time 1D simulations.

In the radioactive tail the same (to an accuracy of $0.5\%$) total $^{56}$Ni
   mass results in a luminosity that is higher by $\approx$3$\%$ compared to
   the early-time 1D models (Fig.~\ref{fig:elts}a).
This behavior of the luminosities is caused by the difference in the expansion
   velocities of the outermost layers for the early and late-time
   1D simulations (Fig.~\ref{fig:velm}).
The higher velocities of the outermost layers for the early-time 1D simulations
   imply a higher rate of work done by radiation pressure.
\citet{Utr_07} showed that the higher the rate of work done by radiation pressure
   is in a SN envelope, the larger is the excess of the gamma-ray luminosity
   (for a given total $^{56}$Ni mass) over the bolometric luminosity in
   the radioactive tail when the SN envelope still remains optically thick
   for gamma rays.

\subsection{Mixing of radioactive $^{56}$Ni and hydrogen}
\label{sec:results-mixing}
%
\begin{table}[t]
\caption[]{Some global properties of mixing in the ejecta.}
\label{tab:mixing}
\centering
\begin{tabular}{@{ } l @{ } c @{ } c @{ } c @{ } c @{ } c @{ } c @{ }}
\hline\hline
\noalign{\smallskip}
 Model & \phantom{e}$\langle v \rangle_\mathrm{Ni}^\mathrm{bulk}$\phantom{e} 
       & \phantom{e}$v_\mathrm{Ni}^{\,\mathrm{bulk}}$\phantom{e}
       & \phantom{e}$\langle v \rangle_\mathrm{Ni}^\mathrm{tail}$\phantom{e}
       & $M_\mathrm{mix}$ & \phantom{m}$\delta M_\mathrm{H}$
       & \phantom{m}$\langle X \rangle$\phantom{m} \\
       & \multicolumn{3}{c}{(km\,s$^{-1}$)}
       & \multicolumn{2}{c}{$(M_{\sun})$} &   \\
\noalign{\smallskip}
\hline
\noalign{\smallskip}
B15-1 &  921 & 3103 & 3241 & 11.45 & 0.111 & 0.040 \\
B15-2 & 1222 & 3355 & 3490 & 11.20 & 0.172 & 0.062 \\
B15-3 & 1807 & 4977 & 5678 & 12.31 & 0.329 & 0.118 \\
N20-P &  924 & 1635 & 1790 &  4.80 & 0.262 & 0.039 \\
N20-C &  930 & 1642 & 1797 &  4.79 & 0.375 & 0.052 \\
W18   &  877 & 1395 & 1472 &  4.10 & 0.062 & 0.011 \\
W20   &  783 & 1374 & 1482 &  5.32 & 0.083 & 0.012 \\
\noalign{\smallskip}
\hline
\end{tabular}
\tablefoot{%
The left column gives the name of the hydrodynamic model.
$\langle v \rangle_\mathrm{Ni}^\mathrm{bulk}$ is the weighted mean velocity
   of the bulk mass of $^{56}$Ni with the $^{56}$Ni mass fraction as weight
   function;
$v_\mathrm{Ni}^{\,\mathrm{bulk}}$, the maximum velocity of the bulk mass of
   $^{56}$Ni;
$\langle v \rangle_\mathrm{Ni}^\mathrm{tail}$, the weighted mean velocity of
   the fast moving $^{56}$Ni tail with the $^{56}$Ni mass fraction as weight
   function;
$M_\mathrm{mix}$, the mass coordinate out to which the bulk mass of $^{56}$Ni
   is mixed; 
$\delta M_\mathrm{H}$, the mass of hydrogen mixed into the He shell; and
$\langle X \rangle$ is the average mass fraction of hydrogen in the mass shell
   defined by $M_\mathrm{CC} \lid m(r) \lid \min(M_\mathrm{He}^{\,\mathrm{core}},
   M_\mathrm{mix})$, where $M_\mathrm{CC}$ and $M_\mathrm{He}^{\,\mathrm{core}}$
   are the mass coordinates given in Tables~\ref{tab:3Dsim} and \ref{tab:presnm},
   respectively.
}
\end{table}
Along with the explosion energy and the total amount of radioactive $^{56}$Ni
   (Table~\ref{tab:3Dsim}), our 3D supernova simulations can be
   characterized by the macroscopic mixing of $^{56}$Ni and hydrogen-rich
   matter occurring during the SN explosion.
In Table~\ref{tab:mixing} we summarize some global properties of the mixing
   induced by the 3D neutrino-driven explosions in the ejecta at
   $t_\mathrm{map}^{\,\mathrm{l}}$, assuming that the bulk mass of
   $^{56}$Ni contains $99\%$ of the total $^{56}$Ni mass and the fast moving
   $^{56}$Ni tail contains the remaining $1\%$.
Dividing the total amount of radioactive $^{56}$Ni into a slow moving bulk
   and a fast moving tail is motivated by the evidence for
   a fast $^{56}$Ni clump of $\sim$10$^{-3}\,M_{\sun}$ observed in SN~1987A
   \citep{UCA_95}, which can be identified with what we define as tail.
Mixing of hydrogen-rich matter may be measured by the mass of hydrogen mixed
   into the He shell or even better by the average mass fraction of hydrogen
   $\langle X \rangle$ in the mass shell defined by
   $M_\mathrm{CC} \lid m(r) \lid \min(M_\mathrm{He}^{\,\mathrm{core}},
   M_\mathrm{mix})$, where $M_\mathrm{CC}$ and $M_\mathrm{He}^{\,\mathrm{core}}$
   are the mass coordinates of the outer edge of the collapsing core and the
   helium core (Tables~\ref{tab:3Dsim} and \ref{tab:presnm}), respectively.

\begin{figure}[t]
   \includegraphics[width=\hsize, clip, trim=18 153 67 210]{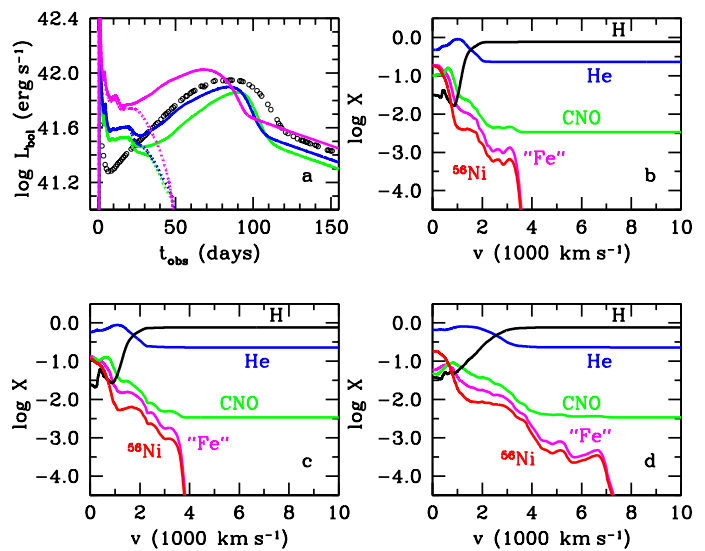}
   \caption{%
   Dependence of the bolometric light curve and the mass fractions at $t=50$ days
      on the explosion energy for the late-time 1D simulations with the
      representative radioactive $^{56}$Ni mass from Table~\ref{tab:3Dsim}.
   Panel \textbf{a} shows the light curves of models B15-1 (\emph{green solid
      line\/}), B15-2 (\emph{blue solid line\/}), and B15-3 (\emph{magenta
      solid line\/}) compared with the observations of SN~1987A obtained by
      \citet{CMM_87, CWF_88} (\emph{open circles\/}).
   The light curves of the corresponding models without radioactive $^{56}$Ni
      are shown with dotted lines.
   Panels \textbf{b}, \textbf{c}, and \textbf{d} show the mass fractions
      as functions of velocity in models B15-1, B15-2, and B15-3,
      respectively.
   The \emph{red line\/} is the mass fraction of radioactive $^{56}$Ni, and
      the \emph{magenta line\/} represents iron-group elements except for
      $^{56}$Ni (see also Fig.~\ref{fig:chcom}).
   }
   \label{fig:depeng}
\end{figure}
Models B15-1, B15-2, and B15-3 show that the production of radioactive
   $^{56}$Ni is proportional to the explosion energy (Table~\ref{tab:3Dsim}).
At the same time, the production of radioactive $^{56}$Ni reveals no clear
   dependence on pre-SN properties, if neutrino-driven
   SN explosion models with similar explosion energies are compared.
The pre-SN models B15, N20, W18, and W20 have different helium cores in
   the mass range from 4 to 7.5\,$M_{\sun}$ (Table~\ref{tab:presnm}),
   different density structures (Fig.~\ref{fig:denmr}),
   different compactness values $\xi_{1.5}$ for 1.5\,$M_{\sun}$ enclosed
   mass (Table~\ref{tab:presnm}), and different chemical compositions
   (Fig.~\ref{fig:chcom}).
A sequence of models B15-2, N20-P, W18, and W20 with comparable explosion
   energies reveals no clear correlation between the production of radioactive
   $^{56}$Ni and these pre-SN properties, although an increasing trend can be
   noticed in this sequence of models with nearly the same
   explosion energy. 

Models B15-1, B15-2, and B15-3 demonstrate that the greater the explosion
   energy (Table~\ref{tab:3Dsim}), the more intense is the mixing of
   radioactive $^{56}$Ni in velocity space (Table~\ref{tab:mixing},
   Figs.~\ref{fig:depeng}b-d), the greater is the average mass
   fraction of hydrogen $\langle X \rangle$, and the higher is the mass of
   hydrogen mixed into the He shell (Table~\ref{tab:mixing}).
The variety of pre-SN properties provides an opportunity to study the
   sensitivity of the outward $^{56}$Ni mixing and the inward hydrogen mixing
   to the structure of the helium core and the He/H composition interface,
   a connection that was discovered by \citet{KPSJM_06} and further studied
   by \citet{WMJ_15}.
It turns out that in the sequence of models B15-2, N20-P, W18, and W20 with
   comparable explosion energies (Table~\ref{tab:mixing}), the mixing of radioactive
   $^{56}$Ni in velocity space and the average mass fraction of hydrogen
   $\langle X \rangle$ decrease, while the mass of hydrogen mixed into
   the He shell does not show any correlation.
There is no leading single parameter that explains this finding, because the
   large-scale mixing is the result of a complex sequence of
   progenitor-structure dependent hydrodynamic instabilities at the composition
   interfaces \citep[cf.][]{WMJ_15}.
It is noteworthy that in all hydrodynamic models the minimum velocity of
   hydrogen-rich matter is less than 100\,km\,s$^{-1}$.

\begin{figure}[t]
   \includegraphics[width=\hsize, clip, trim=18 153 67 210]{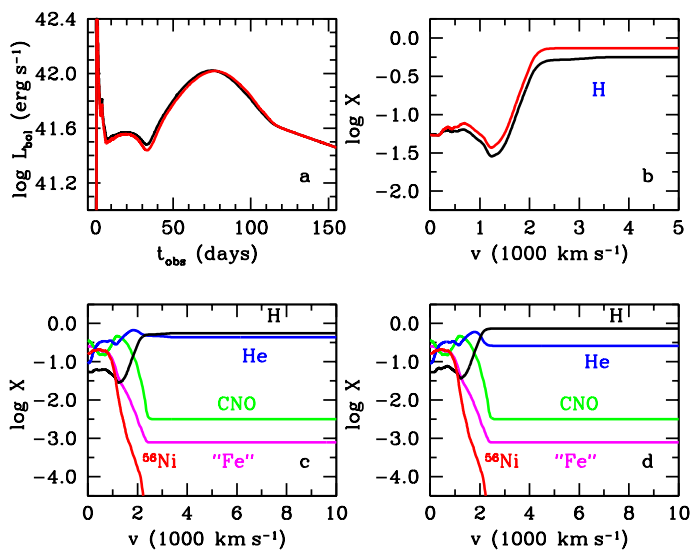}
   \caption{%
   Dependence of the bolometric light curve and the hydrogen mixing on
      the chemical composition of the hydrogen-rich envelope for the
      late-time 1D N20 models with the representative radioactive
      $^{56}$Ni mass from Table~\ref{tab:3Dsim}.
   Panel \textbf{a} shows the bolometric light curves of models N20-P
      (\emph{black line\/}) and N20-C (\emph{red line\/}).
   Panel \textbf{b} shows the mass fraction of hydrogen as function of velocity
      in models N20-P (\emph{black line\/}) and N20-C (\emph{red line\/}).
   Panels \textbf{c} and \textbf{d} give the mass fractions as functions of
      velocity in models N20-P and N20-C, respectively.
   The \emph{red line\/} is the mass fraction of radioactive $^{56}$Ni, and
      the \emph{magenta line\/} represents iron-group elements except for
      $^{56}$Ni (see also Fig.~\ref{fig:chcom}).
   All profiles of the mass fractions are given at $t=50$ days. 
   }
   \label{fig:dephenv}
\end{figure}
From about day 7 to day 30 the bolometric light curve is mainly determined
   by the properties of a cooling and recombination wave (CRW).
In the context of SN~1987A, a basic CRW property reads that the higher the ratio
   of the explosion energy to the ejecta mass, the higher the luminosity in
   the CRW phase because the ejecta expand and cool more quickly
   \citep{Woo_88, Utr_89, SN_90}.
This dependence is distinctly demonstrated by the sequence of models B15-1,
   B15-2, and B15-3 (Table~\ref{tab:3Dsim}, Fig.~\ref{fig:depeng}a).
After the CRW stage, when the radiative diffusion takes place, the radioactive
   decay of $^{56}$Ni and $^{56}$Co nuclides becomes dominant in powering
   the luminosity.
It is evident that in this period the bolometric light curve depends on
   the amount of radioactive material and its distribution over the ejecta.

Finally, we study the influence of the chemical composition of the
   hydrogen-rich envelope on the mixing of radioactive $^{56}$Ni and hydrogen.
For this purpose, we construct an additional model, N20-C, which has the same
   metal composition as model N20-P, except the hydrogen mass fraction
   is enhanced to $X_\mathrm{surf}=0.735$ at the expense of helium in the
   hydrogen-rich envelope (Tables~\ref{tab:presnm} and \ref{tab:3Dsim}).
The increase of the hydrogen abundance from that of the original pre-SN
   model N20-P to the one of model N20-C turns out not to affect the production
   of radioactive $^{56}$Ni and its mixing (Table~\ref{tab:mixing},
   Figs.~\ref{fig:dephenv}c and d).
It only slightly increases the hydrogen mass fraction in the mixing zone
   (Table~\ref{tab:mixing}, Fig.~\ref{fig:dephenv}b).
As a result, the mass fraction of hydrogen in model N20-C exceeds that in
   model N20-P throughout the ejecta.

Another CRW property states that the higher the mass fraction of hydrogen is
   and the lower the mass fraction of helium, the lower is the effective
   temperature in the CRW phase \citep{GN_76, Utr_89}.
For a constant explosion energy, this statement may be reformulated for the
   luminosity in the CRW phase: the higher the mass fraction of hydrogen
   and the lower the mass fraction of helium are, the lower the luminosity.
This is confirmed by the light curves of models N20-P and N20-C for the period
   from day 7 to day 33 (Fig.~\ref{fig:dephenv}a).
A small increase of the mass fraction of hydrogen in the mixing zone makes
   the innermost layers of the ejecta more opaque and, as a consequence, 
   increases the diffusion time and slightly shifts the dome-like maximum of
   the light curve of model N20-C as a whole to later times relative to that
   of model N20-P (Fig.~\ref{fig:dephenv}a).

\subsection{Comparison with observations}
\label{sec:results-cmpobs}
%
\begin{table}[t]
\caption[]{Radioactive $^{56}$Ni masses from fit to observations.}
\label{tab:nifit}
\centering
\begin{tabular}{@{ } l @{ } c @{ } c @{ } c @{ } c @{ } c @{ } c @{ } c @{ } c @{ }}
\hline\hline
\noalign{\smallskip}
 Model & \phantom{e}$M_\mathrm{i}^{\,\mathrm{e}}$
       & \phantom{e}$M_\mathrm{i}^{\,\mathrm{l}}$
       & \phantom{e}$M_\mathrm{f}^{\,\mathrm{e}}$
       & \phantom{e}$M_\mathrm{f}^{\,\mathrm{l}}$
       & $\delta M_\mathrm{Ni}^{\,\mathrm{e}}$
       & $\delta M_\mathrm{Ni}^{\,\mathrm{l}}$
       & $\delta M_\mathrm{tot}^{\,\mathrm{e}}$
       & $\delta M_\mathrm{tot}^{\,\mathrm{l}}$ \\
       & \multicolumn{4}{c}{$(10^{-2}\,M_{\sun})$}
       & \multicolumn{4}{c}{$(10^{-4}\,M_{\sun})$} \\
\noalign{\smallskip}
\hline
\noalign{\smallskip}
B15-1 & 7.52 & 7.29 & 7.42 & 7.26 &  9.2 & 2.4 &  33.2 & 10.2 \\
B15-2 & 7.52 & 7.29 & 7.47 & 7.28 &  4.5 & 0.9 &  24.8 &  6.7 \\
B15-3 & 7.73 & 7.37 & 7.72 & 7.37 &  1.8 & 0.5 &   6.7 &  2.8 \\
N20-P & 7.69 & 7.26 & 7.55 & 7.23 & 14.1 & 2.5 &  97.2 & 21.0 \\
N20-C & 7.69 & 7.26 & 7.55 & 7.23 & 14.1 & 2.5 &  96.4 & 20.1 \\
W18   & 8.53 & 7.33 & 7.68 & 7.26 & 84.8 & 7.2 & 417.3 & 60.8 \\
W20   & 7.87 & 7.31 & 7.46 & 7.24 & 40.5 & 7.1 & 107.7 & 39.6 \\
\noalign{\smallskip}
\hline
\end{tabular}
\tablefoot{%
The names of the hydrodynamic models are given in the left column.
$M_\mathrm{i}^{\,\mathrm{e}}$ and $M_\mathrm{i}^{\,\mathrm{l}}$ are
   the initial $^{56}$Ni masses computed at the onset of light curve modeling
   $t_\mathrm{map}^{\,\mathrm{e}}$ and $t_\mathrm{map}^{\,\mathrm{l}}$,
   respectively (see Table~\ref{tab:3Dsim}).
The other columns provide the $^{56}$Ni mass ejected at day 150, $M_\mathrm{f}$,
   the fallback mass of $^{56}$Ni, $\delta M_\mathrm{Ni}$, during the period from
   $t_\mathrm{map}$ to day 150, and the total fallback mass,
   $\delta M_\mathrm{tot}$, during the same period; all values for the early- and
   late-time 1D simulations.
Note that $M_\mathrm{f}=M_\mathrm{i}-\delta M_\mathrm{Ni}$.
}
\end{table}
A comparison of the calculated light curves of models B15-1, B15-2, and B15-3
   (Table~\ref{tab:3Dsim}) with the observed light curve
   (Fig.~\ref{fig:depeng}a) shows that an explosion energy of
   1.4\,B seems to be close to the optimal value for the given
   pre-SN model.
These light curves were computed with the representative $^{56}$Ni masses from
   Table~\ref{tab:3Dsim}, which are a measure of the $^{56}$Ni production
   during the 3D neutrino-driven explosion.
It is obvious that these representative $^{56}$Ni masses fail to perfectly
   reproduce the observed luminosity in the radioactive tail of SN~1987A
   (Fig.~\ref{fig:depeng}a).

\begin{figure}[t]
   \includegraphics[width=\hsize, clip, trim=18 153 75 321]{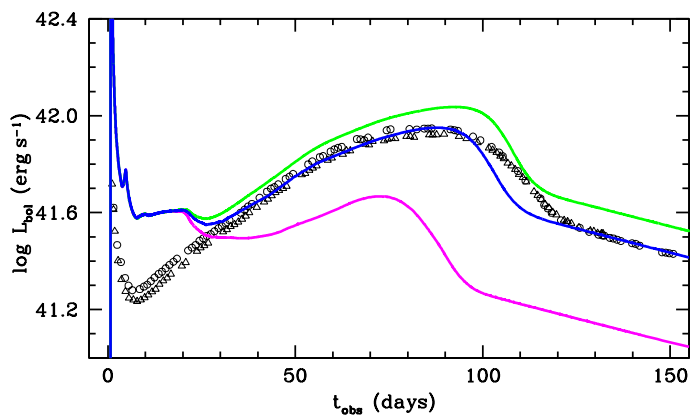}
   \caption{%
   Bolometric light curves of model B15-2 with the mass scaled to fit
      the observed luminosity in the radioactive tail,
      $M_\mathrm{f}^{\,\mathrm{l}}$ (see Table~\ref{tab:nifit})
      (\emph{blue line\/}),
      the mass of radioactive $^{56}$Ni produced directly by our reaction
      network, $M_\mathrm{Ni}^{\,\mathrm{min}}$ (\emph{magenta line\/}), 
      and the aggregate mass of directly produced $^{56}$Ni and tracer nucleus,
      $M_\mathrm{Ni}^{\,\mathrm{max}}$ (\emph{green line\/}),
      (see Table~\ref{tab:3Dsim}) for the late-time 1D simulations
      compared with the observations of SN~1987A obtained by
      \citet{CMM_87, CWF_88} (\emph{open circles\/}) and \citet{HSGM_88}
      (\emph{open triangles\/}).
   }
   \label{fig:nifit}
\end{figure}
To match the observations in the radioactive tail, we adjusted the
   ejected mass of $^{56}$Ni, taking the fallback of radioactive
   $^{56}$Ni after the onset of the light curve modeling
   (Table~\ref{tab:nifit}) into account.
Figure~\ref{fig:nifit} shows the corresponding light curve for model B15-2
   with the adjusted $^{56}$Ni mass.
In addition, the dependence of the light curve of this model on the $^{56}$Ni
   mass and its distribution is demonstrated by the exemplary models with
   minimum, $M_\mathrm{Ni}^{\,\mathrm{min}}$, and maximum,
   $M_\mathrm{Ni}^{\,\mathrm{max}}$, nickel masses (see Table~\ref{tab:3Dsim}).
The uncertainty in the absolute yield of $^{56}$Ni comes, as mentioned above,
   from our small nuclear network and the approximate neutrino transport.
The light curves of the exemplary models reliably comprise the observed
   luminosity in the radioactive tail (Fig.~\ref{fig:nifit}).
Maximum velocities of the bulk of ejected $^{56}$Ni in the models with the
   minimum and maximum $^{56}$Ni masses differ by less than 100\,km\,s$^{-1}$.
This difference is too small to change the luminosity noticeably at the end
   of the CRW stage.
Subsequently, the light curves are completely determined by the total $^{56}$Ni
   mass (Fig.~\ref{fig:nifit}). 

\begin{figure}[t]
   \includegraphics[width=\hsize, clip, trim=18 153 67 215]{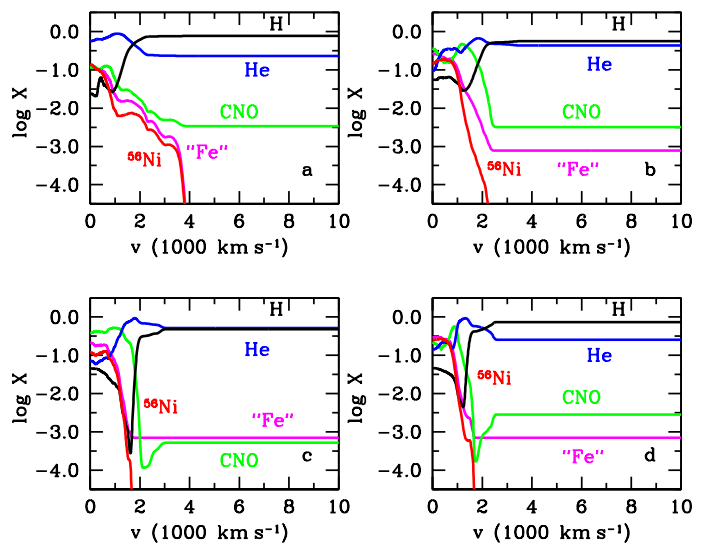}
   \caption{%
   Mass fractions of hydrogen (\emph{black line\/}), helium
      (\emph{blue line\/}), CNO group elements (\emph{green line\/}),
      iron group elements except for $^{56}$Ni (\emph{magenta line\/}),
      and radioactive $^{56}$Ni (\emph{red line\/}) as functions of velocity
      at day 50 in models B15-2 (Panel \textbf{a}),
      N20-P (Panel \textbf{b}), W18 (Panel \textbf{c}),
      and W20 (Panel \textbf{d}) for the late-time 1D simulations.
   In all models the total $^{56}$Ni mass is scaled to fit the observed
       luminosity in the radioactive tail (see Table~\ref{tab:nifit} and
       Fig.~\ref{fig:lcfit}).
   }
   \label{fig:mffit}
\end{figure}
\begin{figure}[t]
   \includegraphics[width=\hsize, clip, trim=18 153 75 321]{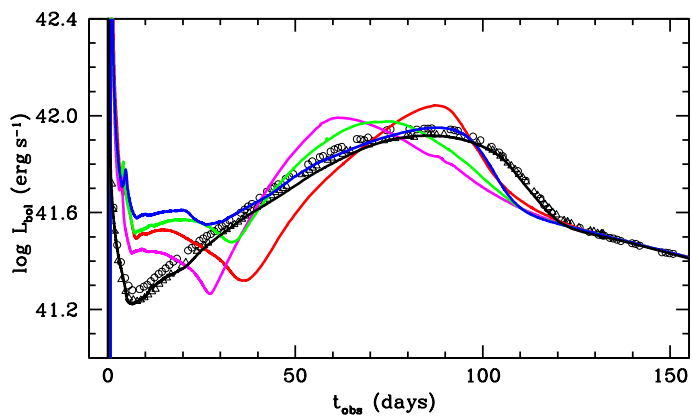}
   \caption{%
   Bolometric light curves of models B15-2 (\emph{blue line\/}),
      N20-P (\emph{green line\/}), W18 (\emph{magenta line\/}), and
      W20 (\emph{red line\/}) for the late-time 1D simulations compared
      with the observations of SN~1987A obtained by \citet{CMM_87, CWF_88}
      (\emph{open circles\/}) and \citet{HSGM_88} (\emph{open triangles\/}). 
   In all models the total $^{56}$Ni mass is scaled to fit the observed
       luminosity in the radioactive tail (see Table~\ref{tab:nifit}).
   In contrast, the \emph{black line} shows the bolometric light curve of
      an optimal model with a nonevolutionary pre-SN
      \citep[][see also Fig.~\ref{fig:denmr}]{Utr_05}.
   }
   \label{fig:lcfit}
\end{figure}

For models B15-2, N20-P, W18, and W20 the resultant mass fraction profiles
   are shown in Fig.~\ref{fig:mffit}, and the fit to the observations
   in the radioactive tail is illustrated by Fig.~\ref{fig:lcfit}.
Because of a different rate of work done by radiation pressure
   (see Sect.~\ref{sec:results-ahexp}) and a different fallback of radioactive
   $^{56}$Ni for the early- and late-time 1D simulations, our estimate of
   the required amount of $^{56}$Ni differs for both runs. 
The initial $^{56}$Ni masses for the early- and
   late-time 1D simulations (Table~\ref{tab:nifit}) fall in between the
   minimum, $M_\mathrm{Ni}^{\,\mathrm{min}}$, and maximum,
   $M_\mathrm{Ni}^{\,\mathrm{max}}$, values (Table~\ref{tab:3Dsim}).
Thus, all 3D neutrino-driven simulations under study are able to synthesize
   the required amount of ejected radioactive $^{56}$Ni.
Note that the fallback of radioactive $^{56}$Ni is not negligible, especially
   in the case of the early-time 1D simulations.

The total mass of radioactive $^{56}$Ni may be evaluated by equating
   the observed bolometric luminosity in the radioactive tail to
   the gamma-ray luminosity.
We call this mass the observed amount of radioactive $^{56}$Ni.
This procedure for SN~1987A gives a mass of $0.0722\,M_{\sun}$.
As expected, all ejected masses of $^{56}$Ni listed in Table~\ref{tab:nifit}
   exceed this value because of the discussed expansion-work effects.
For the early-time 1D simulations, which produce the correct hydrodynamic
   flow in the outermost layers (Fig.~\ref{fig:velm}), these differences are
   on average roughly $5\%$.
We emphasize that the observed amount of radioactive $^{56}$Ni should be
   distinguished from the ejected amount because using the former results in
   an underestimate of the total $^{56}$Ni mass.

An analysis of SN~1987A observations revealed the fact that radioactive
   $^{56}$Ni was mixed in the ejecta up to a velocity of about
   3000\,km\,s$^{-1}$.
Only models B15-1 and B15-2 yield the maximum velocity of the bulk mass of
   $^{56}$Ni consistent with the observations, with $^{56}$Ni being
   concentrated around a velocity of $\sim$1000\,km\,s$^{-1}$
   (Table~\ref{tab:mixing}, Figs.~\ref{fig:depeng}b and c).
Model B15-3 with an explosion energy of 2.59\,B produces mixing that is
   too strong, while models N20-P, N20-C, W18, and W20 with SN~1987A-like
   explosion energies, in turn, yield insufficient mixing.
This weak mixing of $^{56}$Ni in the ejecta of models N20-P, W18, and W20
   slows down the outward diffusion of gamma-rays, admitting a decrease in
   the luminosity after the CRW phase similar to the case of models without
   radioactive $^{56}$Ni (compare Figs.~\ref{fig:depeng}a and \ref{fig:lcfit}).
When the diffusion of gamma-rays becomes efficient, the wide and deep dip of
   the bolometric light curve at around $30-40$ days turns into a luminosity
   rise to maximum (Fig.~\ref{fig:lcfit}).

The existence of a fast $^{56}$Ni clump with an absolute velocity of
   $\approx$4700\,km\,s$^{-1}$ and a mass of $\sim$10$^{-3}\,M_{\sun}$
   in the SN~1987A envelope \citep{UCA_95} is an important observational
   fact that should be explained in the framework of neutrino-driven
   explosions.
The development of clumps is successfully reproduced in 3D neutrino-driven
   simulations as is apparent, for example, from the first panel in the bottom
   row of Fig.~\ref{fig:3D_models}.
The characteristic velocities of the $^{56}$Ni tail, which we have introduced
   to analyze the generated $^{56}$Ni-rich clumps as listed in
   Table~\ref{tab:mixing}, however, fall short of the maximum velocities
   of clumps in SN~1987A.
The only exception is the high-energy model B15-3, whose mean velocity of
   the $^{56}$Ni tail is consistent with the absolute velocity
   of the fast $^{56}$Ni clump in SN~1987A.

Observational evidence for the existence of hydrogen-rich matter within
   the core of heavy elements of SN~1987A implies a deep macroscopic mixing
   down to a velocity of $\sim$500\,km\,s$^{-1}$.
It is remarkable that in all models hydrogen mixing occurring during
   the 3D neutrino-driven simulations results in minimum velocities of
   hydrogen-rich matter of less than 100\,km\,s$^{-1}$
   (Figs.~\ref{fig:depeng}b, c, and d; \ref{fig:dephenv}c and d;
   \ref{fig:mffit}).
This is in good agreement with the spectral observations.

A very disappointing result is that the calculated light curves of all
   models B15-2, N20-P, W18, and W20 roughly having the SN~1987A explosion
   energy are inconsistent with the observations of SN~1987A during
   the first $30-40$ days (Fig.~\ref{fig:lcfit}).
First, in all considered models the initial luminosity peak of the light curve
   is considerably wider than that observed.
This indicates that an adequate configuration of the pre-SN has to be more
   compact.
Second, the calculated light curves of all selected models fail to develop
   a prominent minimum as it is observed at around day 8.
Models without radioactive $^{56}$Ni (Fig.~\ref{fig:depeng}a, dotted lines)
   show that this discrepancy is caused by the structure of the outer layers
   of the pre-SN models (Table~\ref{tab:presnm}, Fig.~\ref{fig:denmr}).
Thus, the structure of the outer layers of the available pre-SN models does not
   agree with that of the BSG Sanduleak $-69^{\circ}202$ just prior to the
   explosion.

In the sequence of models B15-2, N20-P, and W20 all characteristic velocities
   describing mixing of radioactive $^{56}$Ni decrease (Table~\ref{tab:mixing})
   and are inversely related to the times of the distinct local minimum of the
   light curve at which the luminosity starts to grow to the maximum
   (Fig.~\ref{fig:lcfit}).
The higher the characteristic velocity of $^{56}$Ni mixing, the earlier the
   start of the growth of the luminosity to the maximum.
Model W18 differs in this respect because of its low luminosity in the CRW
   phase, which is caused by the lower density in the outer layers of the
   corresponding pre-SN model compared to the other models (Fig.~\ref{fig:denmr}b).
The local minimum of the light curve of model B15-2 is more shallow compared to
   those of models N20-P, W18, and W20.
This reflects the fact that the photosphere enters the mixing zone
   containing the bulk of $^{56}$Ni before the CRW phase ends and, as a
   consequence, the luminosity does not fall off significantly.
In the models N20-P, W18, and W20, the characteristic velocities of
   $^{56}$Ni mixing are so low that a transition from the CRW phase to
   the stage of exhaustion of stored radiation energy occurs earlier than 
   the photosphere enters the mixing zone.
As a result, we see deeper local minima of the light curves
   (Fig.~\ref{fig:lcfit}).
This situation permits us to describe photon diffusion from the mixing zone
   to the photosphere using a simple one-zone approximation
   \citep{Arn_79}.
We estimated the characteristic diffusion times just after the local
   minimum following the CRW phase and found that these times are consistent,
   with an accuracy of $\sim$35$\%$, with the characteristic times of the
   luminosity growth at the same epochs.
For models N20-P, W18, and W20 these times of the luminosity growth are
   about 19, 15, and 19 days, respectively. 

As noted earlier, the broad maximum of the light curve is entirely powered
   by the $^{56}$Ni$\,\rightarrow^{56}$Co$\,\rightarrow^{56}$Fe decay.
The total amount of radioactive $^{56}$Ni is measured by the SN luminosity in
   the radioactive tail, while its mixing throughout the ejecta affects 
   the opacity of the inner layers to which hydrogen makes a major contribution.
The density profile, chemical composition, and opacity of these layers determine
   the optical depth and, consequently, the characteristic diffusion time.
The shape of the bolometric light curve near the maximum, in turn, depends on
   this diffusion time.
The characteristic mass fraction of hydrogen $\langle X \rangle$ of models
   W18 and W20 is about 0.01 (Table~\ref{tab:mixing}), which results in
   a peak-like shape of the light curve maximum.
In contrast, the characteristic mass
   fraction in models B15-2 and N20-P is nearly $0.04-0.06$, which leads to
   a longer diffusion time and a dome-like shape (Fig.~\ref{fig:lcfit}).

It is evident that the dome of the light curve of model B15-2 for the late-time
   1D simulations agrees with the observations of SN~1987A much better than
   that of the other models (Fig.~\ref{fig:lcfit}).
A deficit of the luminosity
   during the transition from the maximum to the radioactive tail, between
   roughly day 100 and day 120, is the only defect in the dome shape.
This luminosity deficit is equivalent to a deficit in the radiated energy.
We have explored the sensitivity of this deficit to different factors and
   found that only an increase of the ejecta mass with a proportional increase
   of the explosion energy (to keep the ratio of ejecta mass to explosion
   energy constant) is able to eliminate this kind of deficit of the luminosity
   after the main maximum.
The explanation is simple: a higher ejecta mass favors an increase of both
   the stored energy and the diffusion time.

Polarimetric data of SN~1987A collected by \citet{Jef_91} show that
   from day 2 to day 20 the $U$ band polarization is about $0.2\%$,
   and the $B, V, R, I$ band polarization is roughly $0.3\%$ with a significant
   scatter.
For our 3D simulations of models B15-2, N20-P, W18, and W20, we approximated the
   asphericity of the outer layers by an ellipsoidal shape at
   $t_\mathrm{map}^{\,\mathrm{l}}$ and found in Sect.~\ref{sec:results-3dtosph}
   that the maximum ratio of the semiaxes is $\approx$1.06. 
According to \citet{Hof_91}, this ratio results in a linear polarization of
   $\approx$0.2$\%$, which is consistent with the broadband polarimetry of
   SN~1987A.

\section{Discussion}
\label{sec:discssn}
%
There are two different approaches to study the peculiar SN~1987A.
In the first approach, the SN event is analyzed performing special evolutionary
   calculations for the progenitor star and subsequent hydrodynamic
   simulations.
It is a long and difficult process from the main-sequence progenitor star
   through its stellar evolution and final explosion to the SN outburst.
The present paper is the first attempt to go the last two steps in the framework
   of neutrino-driven simulations in three dimensions and light curve modeling.
The second approach makes no use of evolutionary progenitor calculations but
   relies on extracting information about the internal structure of the pre-SN,
   the explosion energy, and the amount of radioactive $^{56}$Ni from
   a comprehensive comparison of the extensive observational data with
   adequate hydrodynamic explosion models.
It thus means a reverse engineering process from the observed SN outburst
   to the properties of the pre-SN and the SN explosion.
\citet{Utr_05} carried out this kind of an investigation employing a nonevolutionary
   pre-SN model constructed in an optimal way to reproduce the SN~1987A
   outburst.
This optimal model has a pre-SN radius of $35\,R_{\sun}$, an ejecta mass of
   $18\,M_{\sun}$, an explosion energy of 1.5\,B, and 
   a total $^{56}$Ni mass of $0.0765\,M_{\sun}$.
For a comparison of this optimal model with our selected models B15-2, N20-P,
   W18, and W20, the density profile of the optimal pre-SN star is shown in
   Fig.~\ref{fig:denmr}, and the corresponding light curve in
   Fig.~\ref{fig:lcfit}.

The good agreement between the calculated light curve of the optimal model and
   that observed for the first 40 days confirms our conclusion in
   Sect~\ref{sec:results-cmpobs} that the serious disagreement of the light
   curves of models B15-2, N20-P, W18, and W20 with the observations is caused
   by the large radius of the available pre-SN models and the inappropriate
   structure of their outer layers.
The more compact and different configuration of the optimal pre-SN model
   compared to that of the evolutionary pre-SN models B15, N20, W18, and W20
   (Fig.~\ref{fig:denmr}) implies that all assumptions about the modification
   of convective mixing \citep{WPE_88}, mass loss and convective mixing
   \citep{SNK_88}, and both rotation and convective mixing \citep{WHWL_97}
   have failed to improve the evolutionary calculations of single stars to
   the required extent.
The evolution of the pre-SN in a close binary during a merger with a secondary
   component might offer a more promising possibility to produce the needed
   compact configuration.
It should be noted that hydrodynamic simulations of a binary merger model
   have already explained the formation of the mysterious triple-ring nebula
   surrounding SN~1987A \citep{MP_09}.

The dome-like maximum of the light curve of model B15-2 fits the observations
   of SN~1987A much better than the other models N20-P, W18, and W20 do
   (Fig.~\ref{fig:lcfit}).
It is somewhat surprising that one of the first models for the BSG Sanduleak
   $-69^{\circ}202$ \citep{WPE_88} gives a good result in the framework of
   3D neutrino-driven simulations and light curve modeling, while the more
   advanced pre-SN models W18 and W20 \citep[][respectively]{Woo_07, WHWL_97}
   do less.
This fact shows that the modern modifications in evolutionary calculations of
   single stars do not necessary mean improvements in reproducing the SN~1987A
   progenitor because, for the corresponding structure of the helium core
   and He/H composition interface, 3D neutrino-driven simulations produce
   insufficient amounts of outward $^{56}$Ni mixing and inward hydrogen mixing.

For all models listed in Table~\ref{tab:nifit}, a fit of the calculated light
   curves in the radioactive tail to that observed implies that the total
   ejected mass of $^{56}$Ni, on average, exceeds the observed amount of
   $^{56}$Ni by roughly $5\%$.
In the optimal model of SN~1987A, constructed with a nonevolutionary
   pre-SN model, this excess is nearly $7\%$ \citep{Utr_07}.

\subsection{Assumption of spherical symmetry}
\label{sec:discussion-symmetry}
%
In our analysis of the calculated light curves, we should keep in mind that they
   were obtained with the spherically symmetric radiation hydrodynamics code
   {\sc Crab} after angular averaging of the 3D hydrodynamic flow and
   distribution of chemical elements.
Of course, a more general and more adequate approach would be to solve the
   radiation hydrodynamics equations in 3D for calculating the light curves.
Not applying this 3D radiation hydrodynamics solver, however, only leaves
   the possibility of discussing the possible influence of 3D radiation transfer
   on the light curve.

As shown in Sect~\ref{sec:results-3dtosph}, the outer layers of the ejecta are
   fairly close to a spherically symmetric flow, and the averaging procedure
   introduces small errors associated with the resultant 1D hydrodynamic flow
   and, consequently, with the calculated light curve.
The small asphericity reflects the typical morphology of the hydrogen-rich
   outer layers, while the geometry of radioactive $^{56}$Ni-rich ejecta is
   much more complex (Fig.~\ref{fig:3D_models}, lower row).

The SN ejecta can be divided into the mixing zone characterized by outward
   mixing of $^{56}$Ni and inward mixing of hydrogen, and the outer unmixed
   envelope, which is chemically homogeneous.
The mixing zone contains a number of complex structures made of $^{56}$Ni-rich
   matter rather than a large number of individual clumps.
These $^{56}$Ni structures are distributed almost homogeneously within the
   volume enclosed by a deformed surface, from which a few smaller,
   elongated structures may stick out, depending strongly on the pre-SN 
   structure \citep[c.f. Fig.~14 in][]{WMJ_15}.
Hence, because the angular averaging of the 3D distribution of heavy elements
   transforms macroscopic mixing into microscopic mixing, we need to consider
   the influence of macroscopic mixing of $^{56}$Ni-rich matter and hydrogen
   on the light curve.

The $^{56}$Ni structures can develop into bubbles inflated by radioactive decay
   heating, which may result in a boost of the peak radial velocities of heavy
   elements.
The relatively dense shell forming at the edge of this kind of bubble is
   Rayleigh-Taylor unstable.
After $3-5$ days, the shell breaks into pieces and the $^{56}$Ni-rich
   fragments mix with the surrounding matter \citep{Bas_94}.
Hence, radioactive decay heating enhances macroscopic mixing, tends to
   homogenize matter inside the mixing zone, and may modestly accelerate
   the $^{56}$Ni-rich matter \citep{HB_91, Bas_94}.  

As long as the $^{56}$Ni-rich ejecta remain optically thick for gamma rays,
   the energy from radioactive decay is deposited locally. 
This energy locally powers the thermal and nonthermal emissivity of matter
   which, in turn, forms the SN luminosity, and it tracks the original
   $^{56}$Ni distribution until the optical depth of the $^{56}$Ni-rich ejecta
   approaches unity (in model B15-2, for example, this happens around day 650).
During the subsequent semitransparent phase, diffusion of gamma rays smooths
   the complex morphology of the emissivity making it less pronounced.

\subsection{Viewing angle dependence}
\label{sec:discussion-viewing_angle}
%
For photons the ejecta remain optically thick from the onset of the explosion
   until well into the radioactive tail (in model B15-2, for example, until
   around day 170).
During the CRW phase, up to about day 20 (Fig.~\ref{fig:depeng}a), the SN
   luminosity forms in the hydrogen-rich envelope, which is chemically 
   homogeneous and almost spherically symmetric.
By the end of the CRW phase, the photosphere is approaching the outer edge
   of the mixing zone and then enters the $^{56}$Ni-rich ejecta where 
   gamma rays deposit their energy locally.
From this epoch on, the radioactive decay of $^{56}$Ni and $^{56}$Co
   becomes dominant in powering the luminosity.
As a consequence, the luminosity becomes viewing angle dependent, while
   the ejecta remain optically thick for photons.
Thus, the 3D morphology of thermal and nonthermal emissivity produced by the
   gamma-ray deposition remains unchanged when the photon luminosity
   is sensitive to the heterogeneity of the mixing zone.

From the end of the CRW phase (when in model B15-2, for example, the optical
   depth in the mixing zone is about $10^4$) through the semitransparent phase
   (optical depth of order unity) the photosphere approaches the outer edge
   of the mixing zone, enters it, and eventually disappears.
During this period, two important manifestations of the complex morphology of
   the mixing zone exist.
First, because Thomson scattering by free electrons dominates the opacity,
   diffusing thermal and nonthermal photons cannot provide any local
   information, i.e., the emergent flux depends on the global properties of
   the mixing zone and the SN luminosity becomes viewing angle dependent.

Second, macroscopic inhomogeneities in the chemical composition and in
   the density distribution in the mixing zone reduce the effective opacity
   of matter compared to the homogeneous case, and, consequently,
   the characteristic diffusion time of photons.
The reduced diffusion time results in a luminosity increase after the end of
   the CRW phase.

A closer look at the morphology of the $^{56}$Ni-rich ejecta
   (Fig.~\ref{fig:3D_models}, lower row) suggests that these effects could be
   observationally significant for model B15-2.
In the other models N20-P, W18, and W20 the viewing angle effect might be less
   noticeable because of a closer similarity of the morphology of their
   $^{56}$Ni-rich ejecta to a spherical distribution. 

As a simple illustrative example let us consider an aspherical explosion
   that is strongest and leads to the most intense mixing in one direction
   and is weakest and leads to the least intense mixing in the opposite
   direction.
If we are oriented to the side of the strongest explosion, we can observe
   the light curve with the smooth rising part similar to that of the optimal
   model (Fig.~\ref{fig:lcfit}).
Observations along the opposite direction, in turn, would reveal a local
   minimum in the light curve shortly after the end of the CRW phase like
   that of model B15-2, if in this hemisphere the extent of $^{56}$Ni mixing
   turns out to be insufficient to produce the luminosity required for 
   the smoothness of the light curve.

During both the CRW phase and the late phase when the ejecta are transparent
   for photons, the SN luminosity is independent of the macroscopic properties
   of the mixing zone.
However, between these two phases a viewing angle effect is present.
It first reveals itself when the photosphere is approaching the edge of
   the mixing zone.
The effect then gradually increases as the photosphere enters the mixing zone
   and descends into it.
The growth of the viewing angle effect is caused by the increase of the volume
   of the mixing zone above the photosphere.
After having revealed the morphology of the mixing zone, the viewing angle
   effect approaches a maximum and then starts to decrease, while the mixing
   zone becomes more and more transparent.
The behavior of the viewing angle effect should correlate with the evolution
   of polarization properties.
It is remarkable that broadband polarimetry of SN~1987A shows a similar
   behavior with time \citep{Jef_91}.

\subsection{Ejecta}
\label{sec:discussion-ejecta}
%
The question of the ejecta mass determination is of great importance for any SN.
For SN~1987A, a well-observed and well-studied supernova, this problem
   becomes crucial.
Generally, it would be tempting to determine the ejecta mass by means of
   3D neutrino-driven simulations and light curve modeling based on
   evolutionary pre-SN models.
A relevant evolutionary pre-SN model has to ensure a perfect fit of the
   hydrodynamic model to the observations of SN~1987A.
Unfortunately, the lack of agreement between the calculated light curves and
   the observations of SN~1987A during the first $30-40$ days does not permit
   us to resolve this issue using the available pre-SN models.
In addition, even in the case of model B15-2, having the best dome-like
   maximum of the light curve, there is a deficit of luminosity during
   the transition from the maximum to the radioactive tail in comparison with
   the observed light curve (Fig.~\ref{fig:lcfit}).

It is noteworthy that among a family of SN~1987A-like events with dome-like
   light curves there is a peculiar object, SN~2000cb.
The latter is characterized by an explosion energy of 4.4\,B
   and mixing of radioactive $^{56}$Ni up to a velocity of 8400\,km\,s$^{-1}$
   \citep{UC_11}.
SN~1987A and SN~2000cb show that the greater the explosion energy is, the higher
   is the degree of $^{56}$Ni mixing.
The same trend is demonstrated by models B15-1, B15-2, and B15-3
   (Tables~\ref{tab:3Dsim} and \ref{tab:mixing}).
In the case of model B15-3 the maximum velocity of $^{56}$Ni matter is about
   7300\,km\,s$^{-1}$ (Fig.~\ref{fig:depeng}d), which is comparable with that
   of SN~2000cb.
Unfortunately, a more detailed and meaningful assessment of this SN seems to be
   questionable in the framework of neutrino-driven explosions because the
   explosion energy of 4.4\,B for SN~2000cb may be hard to reach with explosions
   driven by the neutrino-heating mechanism \citep{UJM_12}. 

The important observation of a high-velocity clump of $\sim$10$^{-3}\,M_{\sun}$
   with a speed of about 4700\,km\,s$^{-1}$ in the envelope of SN~1987A
   \citep{UCA_95} is not reproduced by any of our investigated models with
   SN~1987A-like explosion energies (see Sect.~\ref{sec:results-cmpobs} and
   Table~\ref{tab:mixing}), also, in particular, not by model B15-2.
Also, a modest increase of the explosion energy does not remove this discrepancy
   because the maximum $^{56}$Ni velocities scale roughly and moderately with
   the square root of the explosion energy \citep{WMJ_15}. 

\subsection{Progenitors}
\label{sec:discussion-progenitor}
%
Comparing the observational data of the BSG Sanduleak $-69^{\circ}202$ and
   evolutionary stellar models, \citet{SNK_88} and \citet{Woo_88}
   concluded that the star, at the time of its explosion, had a helium-core
   mass of $\approx$6\,$M_{\sun}$.
From an inspection of Table~\ref{tab:presnm}, it is evident that only pre-SN
   models N20 and W20 have helium-core masses consistent with the above estimate.
Unfortunately, using these models in our 3D neutrino-driven simulations results
   in mixing of $^{56}$Ni that is too weak and, as a consequence, in a disagreement of
   the calculated light curves with the observations of SN~1987A
   (Fig.~\ref{fig:lcfit}).
Thus, a correct mass of the helium core in the pre-SN configuration does not
   guarantee an adequate explanation of the SN~1987A phenomenon.
In this context, a helium-core mass of $4.08\,M_{\sun}$ in the pre-SN model B15
   appears as a serious shortcoming to fit the observational data of
   the BSG Sanduleak $-69^{\circ}202$.
This conclusion seems in line with our finding (Sect.~\ref{sec:results-cmpobs})
   that the deficit of model B15-2 in matching the late-time light curve of
   SN~1987A can only be cured by an increase of the ejecta mass and
   a proportional increase of the explosion energy.

The 3D simulations we employed started from
   spherically symmetric progenitor models.
To initiate the growth of nonradial hydrodynamic instabilities,
   small seed perturbations had been imposed as zone-to-zone variations of
   the radial velocity with chosen amplitude in the whole computational
   domain at the start of the 3D simulations shortly after core bounce.
However, \citet[][see also references in their paper]{AM_11} pointed out
   that turbulent convective shell burning might produce large-scale
   precollapse asphericity in the progenitor core, and, in particular, of
   the O-burning shell, which could affect explosion asymmetries,
   neutron-star kicks, and nickel mixing during the explosion.
In addition, \citet{CO_13} showed that the revival of the stalled SN shock
   by neutrino heating could be triggered by nonradial flows in the progenitor
   core because of perturbations in the Si/O burning shells, when advected
   through the accretion shock, enhance nonradial mass motions and turbulent
   pressure in the postshock region \citep{CO_15, MJ_15}.
While these effects might be of relevance in the context discussed in our paper,
   it is unclear whether the relatively large initial perturbations that are
   needed to make a sizable impact are realistic \citep{MJ_15}.
A quantitatively meaningful assessment is currently not possible because the
   perturbation pattern and amplitude in 3D stars prior to core collapse must
   still be determined by multidimensional simulations of the late stages of
   stellar evolution.

\subsection{Numerical resolution}
\label{sec:discussion-numerics}
%
It is unlikely that the inability of our models to reproduce the high-velocity
   clump observed in the envelope of SN~1987A
   (see Sect.~\ref{sec:discussion-ejecta}) is a fundamental problem of the
   neutrino-driven explosion mechanism.
A number of possibilities may naturally cure this deficit.

Besides possible large-scale progenitor asphericities prior to stellar core
   collapse (see the discussion in Sect.~\ref{sec:discussion-progenitor})
   and a different progenitor structure, which plays a crucial role for
   the long-time propagation
   and deceleration of the iron-group ejecta \citep[cf.][]{WMJ_15, HJM_10},
   the most important aspect is the numerical resolution of the 3D simulations.
For reasons of computational efficiency, the employed 3D models were computed
   with an angular resolution of only two degrees.
Comparisons with one-degree simulations showed good agreement in the
   distribution and motion of the bulk of nickel, but the amount and speed of
   the fastest radioactive $^{56}$Ni turned out to be highly sensitive to
   the grid resolution.
Both the amount and speed of the fastest $^{56}$Ni decrease with time as
   the flow expands over the Eulerian grid \citep{WMJ_15}.
These effects are undesirable artificial, but unavoidable consequences of
   numerical viscosity and diffusivity, which become more serious for coarser
   resolution.
In a 3D explosion simulation of our considered B15 progenitor with similar
   explosion energy, \citet{HJM_10} obtained maximum $^{56}$Ni velocities of
   up to 4500\,km\,s$^{-1}$, which is in the ballpark of the fastest
   radioactive material observed in SN~1987A.
This simulation employed a regular spherical polar grid and thus had higher
   spatial (linear) resolution in the azimuthal direction near the polar axis than
   the best models of \citet{WMJ_15} with their Yin-Yang grid.
The result of \citet{HJM_10} demonstrates that it is well possible to retain
   nickel at speeds considerably above 4000\,km\,s$^{-1}$ in the B15 model
   with explosion energies around $1-1.5$\,B.

\section{Conclusions}
\label{sec:conclsns}
%
Comparing results of 3D neutrino-driven simulations and light curve modeling
   with the observations of SN~1987A allows us to draw the following principal
   conclusions, which, of course, depend on the considered set of pre-SN
   models:
\begin{itemize}
\item[$\bullet$]
The dome of the light curve of model B15-2 for the late-time 1D simulation
   agrees with the observations of SN~1987A much better than that of the other
   models N20-P, W18, and W20 (Fig.~\ref{fig:lcfit}).
There is only one defect in the dome shape -- a deficit of the luminosity
   during the transition from the maximum to the radioactive tail.
\item[$\bullet$]
To fit the observations in the radioactive tail, we adjusted the
   ejected amount of $^{56}$Ni, taking fallback of radioactive
   $^{56}$Ni starting from the onset of light curve modeling into account
   (Table~\ref{tab:nifit}).
The required initial $^{56}$Ni masses in the early- and late-time 1D simulations
   fall in between the minimum and maximum estimates obtained in our 3D
   explosion models, implying that all 3D neutrino-driven simulations under
   study are able to synthesize the ejected amount of radioactive $^{56}$Ni.
\item[$\bullet$]
The analysis of SN~1987A observations revealed the fact that radioactive
   $^{56}$Ni was mixed in the ejecta up to a velocity of about
   3000\,km\,s$^{-1}$.
Only models B15-1 and B15-2 yield a maximum velocity of the bulk of $^{56}$Ni
   consistent with the observations, with $^{56}$Ni being concentrated around
   a velocity of about 1000\,km\,s$^{-1}$
   (Table~\ref{tab:mixing}, Figs.~\ref{fig:depeng}b and c).
\item[$\bullet$]
In hydrodynamic models W18 and W20, fallback of the radioactive $^{56}$Ni
   is not negligible (Table~\ref{tab:nifit}).
\item[$\bullet$]
Our set of pre-SN models provides the opportunity to study the sensitivity
   of outward $^{56}$Ni mixing and inward hydrogen mixing to 
   the structure of the helium core and the He/H composition interface.
It turned out that mixing of radioactive $^{56}$Ni in velocity space
   and the average mass fraction of mixed hydrogen $\langle X \rangle$ decreased,
   while the mass of hydrogen mixed into the He shell did not show any
   systematics in the sequence of models B15-2, N20-P, W18, and W20 with
   comparable explosion energies (Table~\ref{tab:mixing}).
\item[$\bullet$]
It is remarkable that in all models hydrogen mixing, occurring during the
   3D neutrino-driven simulations, results in minimum velocities of
   hydrogen-rich matter of less than 100\,km\,s$^{-1}$, which is in a good
   agreement with the spectral observations of SN~1987A.
\end{itemize}

In addition, we can summarize a few secondary findings:
\begin{itemize}
\item
The influence of the chemical composition of the hydrogen-rich envelope on
   the mixing of radioactive $^{56}$Ni and hydrogen was studied with
   model N20-P ($X_\mathrm{surf}=0.56$) and the additional model N20-C
   ($X_\mathrm{surf}=0.735$).
This increase of the hydrogen abundance from that of the pre-SN to the
   LMC-typical value does not affect the mixing of radioactive $^{56}$Ni and
   only slightly increases the hydrogen mass fraction in the mixing zone
   (Table~\ref{tab:mixing}).
\item
Mixing between $t_\mathrm{map}^{\,\mathrm{e}}$, prior to shock breakout, and
   $t_\mathrm{map}^{\,\mathrm{l}}$, long after shock breakout when the ejecta
   expand almost homologously, increases both the mass fraction of hydrogen
   in the metal core and that of radioactive $^{56}$Ni in the outer part of
   the mixing zone, with the maximum $^{56}$Ni velocity remaining unchanged
   (Fig.~\ref{fig:elts}).
\item
Models B15-1, B15-2, and B15-3 demonstrate that the greater the explosion
   energy (Table~\ref{tab:3Dsim}), the more intense is the mixing of
   radioactive $^{56}$Ni in velocity space, the greater is the average mass
   fraction of mixed hydrogen $\langle X \rangle$, and the higher is the mass of
   hydrogen mixed into the He shell (Table~\ref{tab:mixing}).
\end{itemize}

We conclude with some critical comments about the pre-SN models for
   SN~1987A, which we used in our 3D neutrino-driven simulations and light curve
   modeling.
Common shortcomings of all pre-SN models studied are a pre-SN radius that
   is too large to fit the initial luminosity peak of SN~1987A and a structure
   of the outer layers that is inadequate to match the observed light curve
   during the first 40 days.
At the same time, the structure of the helium core and of the He/H composition
   interface in the pre-SN model B15 facilitates the production of a sufficient
   amount of outward $^{56}$Ni mixing and of inward hydrogen mixing in model
   B15-2 and, as a consequence, results in a dome-like maximum of the light
   curve that fits the observations much better than the maxima of the other
   models do.
Unfortunately, a helium-core mass of $4.08\,M_{\sun}$ in the pre-SN model B15
   is in conflict with observational data of the BSG Sanduleak $-69^{\circ}202$.
Therefore, there is no evolutionary pre-SN model with the required helium-core
   mass of $6\,M_{\sun}$ and a suitable structure of the helium core and He/H
   composition interface.
The lack of an adequate pre-SN model for the well-observed and well-studied
   SN~1987A is a real and pressing challenge for the stellar evolution theory
   of massive stars.
An important lesson from our study of SN~1987A is that the evolutionary
   picture for the late stages of massive stars may require revision in ways
   yet to be determined.

Finally, we state that the available pre-SN models, self-consistent
   3D neutrino-driven simulations, and light curve modeling can explain
   the basic observational data of SN~1987A, except for those related to
   the detailed pre-SN structure of the outer layers, within the paradigm
   of the neutrino-driven explosion mechanism.

\begin{acknowledgements}
%
We would like to thank Ken Nomoto and Stan Woosley for providing us with
   the pre-SN data.
V.P.U. was supported by the guest program of the Max-Planck-Institut f\"ur
   Astrophysik.
At Garching, funding by the Deutsche Forschungsgemeinschaft through grants
   SFB/TR7 ``Gravitational Wave Astronomy'' and EXC 153 ``Origin and Structure
   of the Universe'' and by the EU through ERC-AdG No.\ 341157-COCO2CASA
   is acknowledged.
Computation of the 3D models and postprocessing of the data were done on Hydra
   of the Rechenzentrum Garching.
%
\end{acknowledgements}


\end{document}